\documentclass{aa}  

\usepackage{subfigure}
\usepackage{graphicx}
\usepackage[para,online,flushleft]{threeparttable}
\usepackage{placeins}
\usepackage{txfonts}
\begin{document}

\title{A proper motion catalogue for the Milky Way's nuclear stellar disc}

\author{B. Shahzamanian\inst{1}, R. Sch\"odel\inst{1}, F. Nogueras-Lara\inst{2}, A. Mart\'i­nez-Arranz\inst{1}, Mattia C. Sormani\inst{3}, A. T. Gallego-Calvente\inst{1}, E. Gallego-Cano\inst{1}, A. Alburai\inst{1}}

   \institute{Instituto de Astrof\'isica de Andaluc\'ia (CSIC), Glorieta de la Astronom\'i­a s/n, 18008 Granada, Spain\\
       \email{shahzaman@iaa.es}
          \and
             Max-Planck-Institut f\"ur Astronomie, K\"onigstuhl 17, 69117 Heidelberg, Germany
          \and
            Universit\"at Heidelberg, Zentrum f\"ur Astronomie, Institut f\"ur theoretische Astrophysik, Albert-Ueberle-Str. 2, 69120 Heidelberg, Germany
  }

\date{Received:/ Accepted:  }


\abstract {We present the results of a large-scale proper motion study of the central $\sim 36' \times 16'$ of the Milky Way, based on our high angular resolution GALACTICNUCLEUS survey (epoch 2015) combined with the HST Paschen-$\alpha$ survey (epoch 2008). Our catalogue contains roughly $80,000$ stars, an unprecedented kinematic dataset for this region. We describe the data analysis and the preparation of the proper motion catalogue. We verify the catalogue by comparing our results with measurements from previous work and data. We provide a preliminary analysis of the kinematics of the studied region. Foreground stars in the Galactic disc can be easily identified via their low reddening. Consistent with previous work and with our expectations, we find that stars in the nuclear stellar disc have a smaller velocity dispersion than inner bulge stars, in particular in the direction perpendicular to the Galactic plane. The rotation of the nuclear stellar disc can be clearly seen in the proper motions parallel to the Galactic plane. Stars on the near side of the nuclear stellar disc are less reddened than stars on its far side. Proper motions  enable us to detect co-moving groups of stars that may be associated with young clusters dissolving in the Galactic centre that are difficult to detect by other means. We demonstrate a technique based on a density clustering algorithm that can be used to find such groups of stars. 
}


\authorrunning{B. Shahzamanian} 
\titlerunning{Galactic centre proper motions}
\maketitle
\section{Introduction}
\label{section:Introduction}

The Galactic centre (GC) is a benchmark object for studying the galactic nuclei of present-day large spiral galaxies. It contains the super-massive black hole Sagittarius\,A* \citep[Sgr\,A*; e.g.][]{Eckart:1996tg,Eckart:1997jl,Ghez:2008fk}, the nearest nuclear star cluster (NSC), and the nearest nuclear stellar disc \citep[NSD; see][]{Launhardt:2002nx,Schodel:2014bn}. The GC is located only about 8\,kpc from Earth and is therefore the only nucleus of a galaxy in which we can observationally resolve a significant fraction of stars. 

The NSD is a dense rotating stellar structure with a radius of roughly 200\,pc and a scale height of $\sim$50\,pc \citep{Launhardt:2002nx,Nishiyama:2013uq,Schoenrich:2015,gallego-cano:2020}. The star formation history of the NSD was thought until recently to be characterised by quasi-continuous, quasi-constant star formation \citep{Figer:2004fk}. However, \cite{nogueras:2020} showed that the luminosity functions based on data from the new GALACTICNUCLEUS survey \citep[GNS;][]{nogueras:2019cat, nogueras:2019ext} are inconsistent with this scenario. Instead, about 90\% of the stars in the NSD are older than 8 Gyr, and there was a pronounced star formation event about 1 Gyr ago that contributed 5\% of the mass of the NSD.
 Based on the K-band Multi Object Spectrograph (KMOS)/Very Large Telescope (VLT) data of \citet{fritz:2020}, \cite{schultheis2021} recently demonstrated that the NSD is kinematically and chemically distinct from both the Galactic bar and the NSC. They propose that the metal-rich stars in the NSD may have formed from gas in the central molecular zone.

\begin{figure*}[!t]
  \includegraphics[width=1.\textwidth]{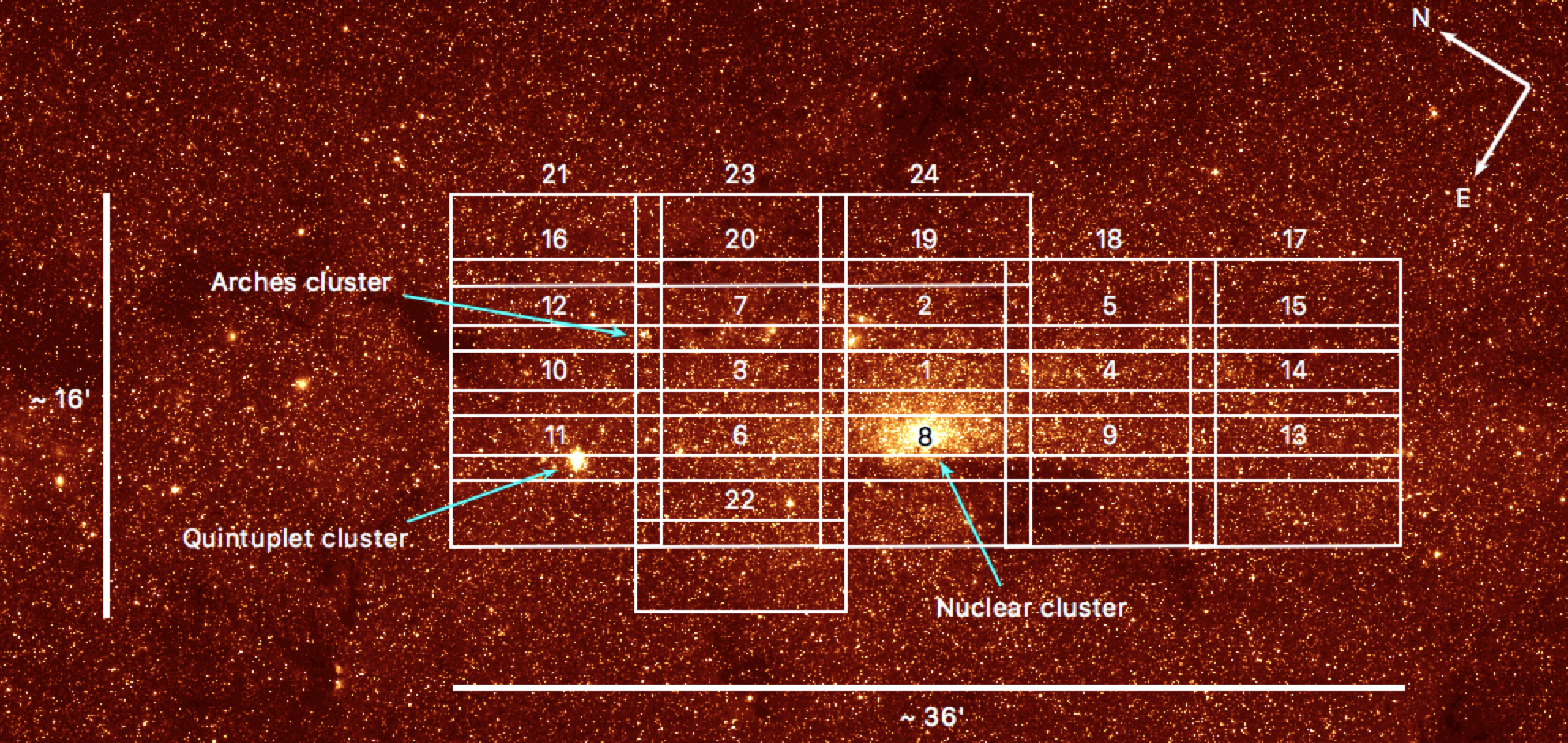}
  \caption{Pointings of the central region of the GNS used in this work overlaid on a 4.5 $\mu$m Spitzer/IRAC extinction-corrected image of the GC from \cite{schoedel2014}. Each white rectangle shows a field with a size of $\sim 7.95' \times 3.43'$ and is identified by a number. The NSC, the Quintuplet, and the Arches clusters are also marked.
 }

  \label{fig:GNS} 
\end{figure*}


Averaged by volume, the GC has been the Milky Way's most active star forming site over the past $\sim30$\,Myr. Three of the most massive young star clusters in the Milky Way are located in the NSD, the Arches, the Quintuplet, and the Central Parsec clusters. They have ages of 3-6\,Myr and total masses $\gtrsim10^{4}$\,M$_{\odot}$ \citep[e.g.][]{Figer:2004fk}. 
The recent star formation history of the NSD implies that there should be at least ten more massive young clusters in the NSD that have still not been discovered \citep{Matsunaga:2011uq,nogueras:2020}. The most likely explanation for the missing clusters is that tidal disruption makes their density drop below the high background density of old stars within < 10\,Myr  \citep{Portegies-Zwart:2002fk, Kruijssen2014}. Due to the high and variable extinction towards the GC, these clusters cannot be discovered by means of colour magnitude diagrams (CMDs) with currently available data \citep{nogueras:2021}. 

A reasonably complete picture of the stellar population at the GC is difficult to obtain because of (1) the large and highly variable interstellar extinction towards the GC \citep{schoedel2010, Fritz:2011fk, nogueras:2019ext} that makes it extremely challenging to classify stars based on their near-infrared colours; and (2) overlapping and co-penetrating Galactic components (the Galactic disc and bulge, NSD, NSC, and young clusters) along the line-of-sight.

Proper motions can be a powerful tool for disentangling different kinematic components in complex regions of the Milky Way, such as the NSD and inner bulge. In particular, in the GC, proper motions can help us assign stellar populations to the bulge or the NSD \citep[see e.g. the discussion in][]{Matsunaga2018}. With proper motions we may also be able to identify stellar streams associated with past accretion events \citep[e.g.][]{Feldmeier-Krause:2020pi} and to find young stellar clusters in the form of compact co-moving groups \citep[e.g.][]{Stolte:2008uq,Hosek:2015sh,Rui:2019ch,Shahzamanian:2019}.

So far, precise proper motions in the GC have generally been measured on fields smaller than a few square arcminutes \citep[e.g.\ to study the Central Parsec, Arches, or Quintuplet clusters;][]{Hosek:2015sh,Rui:2019ch,Trippe:2008it,Schodel:2009zr}. The recent work by \cite{Libralato:2021} covers a significantly larger field of about 56 arcmin$^{2}$ within the NSD (and a similarly large area outside of it) with the Wide-Field Camera 3 (WFC3)/infrared (IR) and Advanced Camera for Surveys (ACS)/Wide-Field Channel (WFC) of the Hubble Space Telescope (HST). However, their work does not cover the innermost $\sim$100\,pc of the NSD.
Gaia is blind towards the GC but the Gaia Data Release 2 \citep[DR2;][]{gaia:2016, gaia:2018} catalogue provides precise proper motions of foreground stars that can be used to determine absolute astrometry of GC sources. The Vista Variables in the Via Lactea (VVV) survey \citep{Minniti:2010fk} covers the entire inner Galaxy but is severely limited by seeing-limited resolution and saturation in the GC. Therefore, precise proper motions from the VVV in the NSD are only available for stars in a very limited magnitude range \citep[approximately $12\lesssim K_{s}\lesssim 13.5$;][]{Smith:2018qf}. 

The GNS has been designed specifically for the GC and stands out from other surveys due to its  uniform $0.2"$ angular resolution, which minimises confusion and maximises astrometric precision. Confusion is more than a factor of 10 lower in the GNS, and its dynamic range is about five magnitudes higher than in seeing-limited surveys of the GC, such as the VVV \citep{nogueras:2019cat}. It is therefore the ideal basis for a proper motion survey of the NSD. Currently, the only other dataset that approximately matches the GNS in angular resolution and observed field is the the HST  Paschen-$\alpha$ survey \citep[H-P$\alpha$S;][]{Wang:2010fk, Dong:2011ff}. In this work we combine the 2008 H-P$\alpha$S with the 2015 GNS to obtain the first proper motion catalogue for the central $\sim 36' \times 16'$ of the NSD. We have already demonstrated this method  on a small field and identified a new co-moving group in \citet{Shahzamanian:2019}. This work presents the proper motions for the entire overlapping field of the GNS and H-P$\alpha$S.

This paper is structured as follows. In Sect.~\ref{section:Data sets} we describe the data and methodology. Section~\ref{section:Proper motions} describes the proper motion catalogue and compares it with other data in overlapping fields. We analyse the kinematics of stellar populations at the GC in Sect.~\ref{section:Kinematics in the GC}. In Sect.~\ref{section:Finding co-moving groups} we describe a clustering search algorithm and its application. We summarise our results in Sect.~\ref{section:summary}.

\section{Data and methodology}
\label{section:Data sets}

We use two datasets, the GNS \citep{nogueras:2019cat}, epoch 2015/2016, and the H-P$\alpha$S \citep{Wang:2010fk, Dong:2011ff}, epoch 2008, with a time baseline of seven to eight years between them. Here we only briefly summarise our methodology. Further details are described in \citet{Shahzamanian:2019}.


\begin{figure}
  \includegraphics[width=\columnwidth]{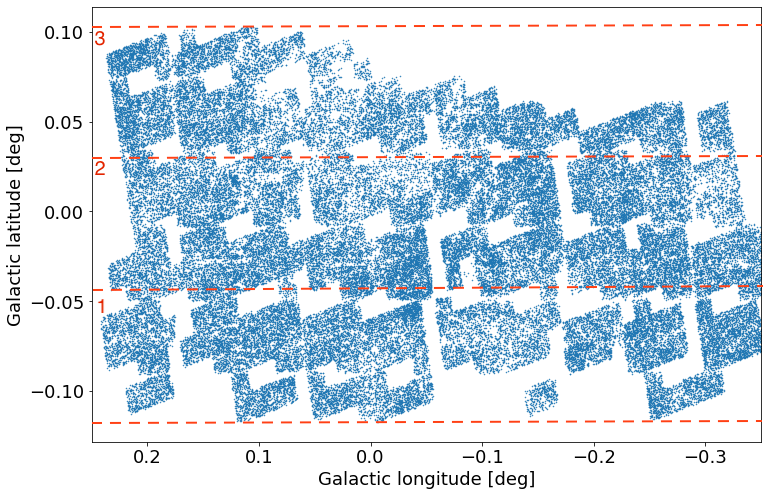}
  \caption{Positions of the stars in our proper motion catalogue. Three sub-regions of the catalogue from the vertical division are also shown (see Sect. 4.1).}
 \label{fig:measured_pos}
\end{figure}


\begin{figure}[!htb]
  \includegraphics[width=\columnwidth]{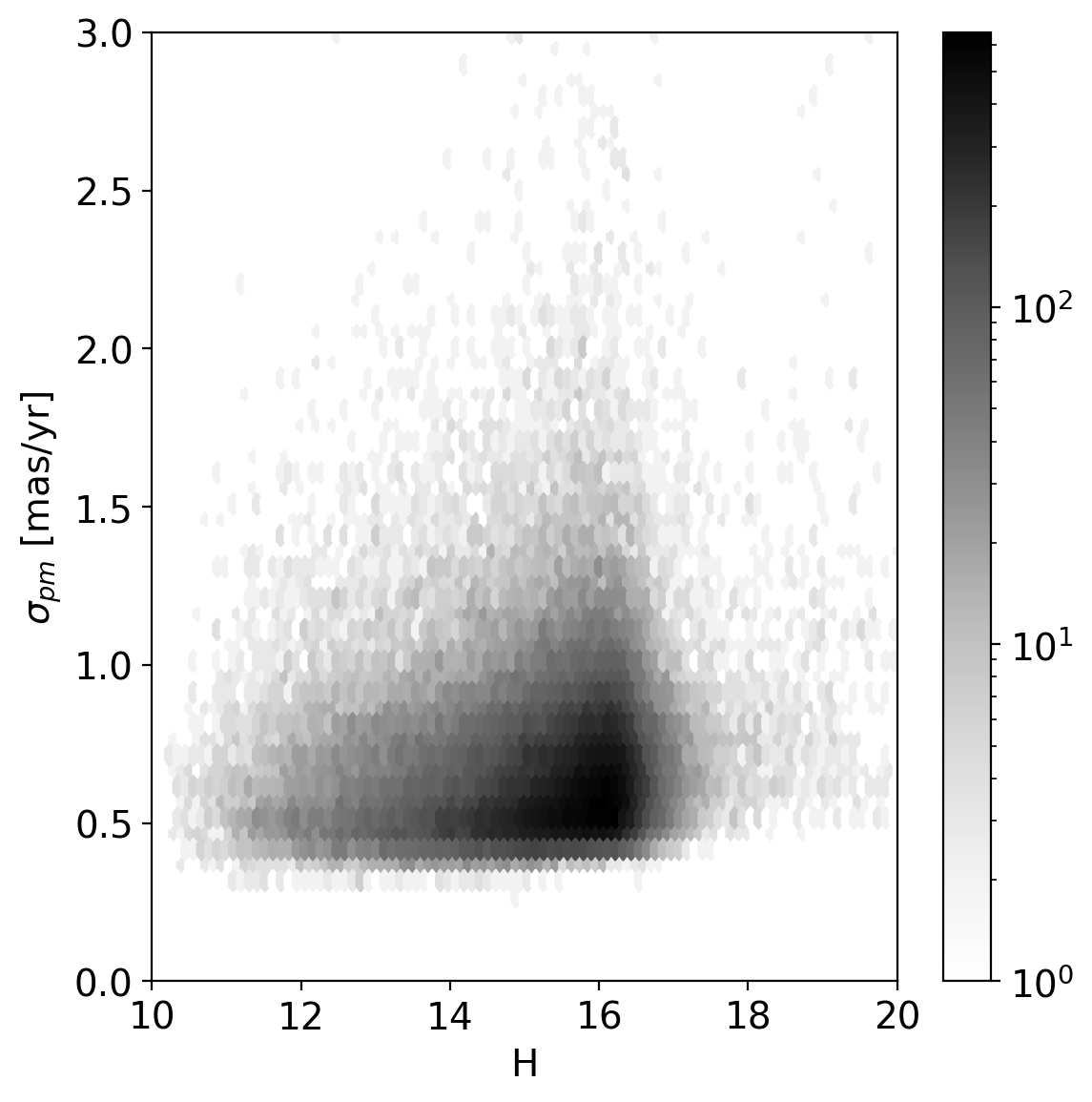}
  \caption{Uncertainties of proper motions in the catalogue as a function of H magnitude (see Sect.~\ref{subsection:Proper motions catalogue}).}
  \label{fig:dpm}
\end{figure}

All relevant details of the GNS are described in \cite{Nogueras2018a,nogueras:2019cat}. \citet{Dong:2011ff} describes all relevant information about the H-P$\alpha$S. Both surveys have very similar angular resolutions of $\sim$$0.2"$. The GNS observations were taken in the near-infrared ($J$, $H$, and $K_{\rm s}$) with the High Acuity Wide-field K-band Imager (HAWK-I) at the VLT. HAWK-I has a detector composed of four different chips with a gap in between them \citep{Kissler-Patig:2008uq}. For the proper motion catalogue, we used only the $H$-band data of the GNS because the extreme extinction in $J$ ($A_{J} \gtrsim7$) implies a significantly smaller number of detections than at longer wavelengths and because saturation is a concern in $K_{\rm s}$ for stars brighter than $K_{\rm s} \sim 11$. We only selected GNS stars with a relative astrometric uncertainty of $<2$\,milliarcseconds (mas) along each axis \cite[see Fig.~A.1 of][]{Shahzamanian:2019}. The central region of the GNS includes 30 pointings and is centred on Sgr~A*, covering the GC area of about $36' \times 16'$. We analysed 24 pointings (labelled 1-24) of the central region that have full or partial overlap with  the H-P$\alpha$S (see Fig.~\ref{fig:GNS}).

The H-P$\alpha$S maps an area of $\sim 36' \times 16'$ that has been acquired with the Near-Infrared Camera and MultiObject Spectrometer (NICMOS) Camera 3 (NIC3) on the HST. The field-of-view (FoV) of NIC3 is $51.2'' \times 51.2''$, and the final mosaic has an angular resolution of $0.2''$ and a pixel size of $0.1''$. Its FoV per pointing is about 37 times smaller than a single GNS pointing. The survey and the source list catalogue are described in \cite{Dong:2011ff}. Because of the mosaicing process, the astrometric positions given in this list have uncertainties of a few tens of mas and can therefore not be directly used to measure proper motions. We therefore carried out astro-photometry on each survey image with the \textit{StarFinder} point spread function (PSF) fitting  package \citep{Diolaiti:2000fk}. As uncertainties we used the formal uncertainties provided by \textit{StarFinder}. We used the data acquired with the narrow-band filter F190N and only accepted stars with less than 3~mas astrometric uncertainty along each axis for our proper motion analysis \cite[see Fig.~A.2 of][]{Shahzamanian:2019}. All these stars have a significantly smaller positional uncertainty in the HAWK-I images.

The GNS data were reduced and calibrated for each chip and pointing individually. Each GNS chip overlaps (fully or partially) with a different number of NICMOS/HST images. Assuming net zero motion and rotation for the stars present in each image, we transformed the stellar positions of the H-P$\alpha$S images into the reference frame of the corresponding GNS pointing and chip via a third-order polynomial transformation, as described in \citet{Shahzamanian:2019} and  \citet{schoedel2009}. We calculated the alignment uncertainties with the jackknife method,\ repeating the procedure multiple times and dropping different groups of reference stars in each repetition. The alignment uncertainties mainly depend on the number and distribution of reference stars for this procedure, which are determined primarily by the size  of the overlapping region of H-P$\alpha$S and GNS pointings and by the stellar surface density.

After alignment, we identified common stars by the condition that they had to coincide within a  $0.1''$ radius (i.e.\ half the resolution limit of the two surveys). Since a proper motion of 1\,mas\,yr$^{-1}$ corresponds to a physical velocity of about 40\,km\,s$^{-1}$ at the distance of the GC \citep[we assumed 8\,kpc; see e.g.][]{Contreras-Ramos:2018kf,Abuter:2019fk,Do:2019ha}, our criterion corresponds to a maximum detectable velocity of $100$\,mas$/7$\,yr\,$\approx14$\,mas\,yr$^{-1}$ or $600$\,km\,s$^{-1}$. This is roughly five times larger than the maximum velocity dispersion expected for the different populations in the target field, which is $\lesssim3$\,mas\,yr$^{-1}$ or 120\,km\,s$^{-1}$ \citep[e.g. Field Gaussian\,2 of the multi-component fit of][]{Rui:2019ch}. 

Finally, we obtained the proper motions for each star parallel and perpendicular to the Galactic plane, $\mu_{l}$ and $\mu_{b}$, by dividing their displacement by the time baseline of seven years (eight years for the two GNS fields observed in 2016). We obtained the uncertainties of the proper motions by quadratically combining the uncertainties of the astrometric positions in the two epochs with the alignment uncertainty between the epochs. The latter depends on the position of any given star within each NICMOS/HST image \citep[see][]{Shahzamanian:2019}.

Because of the overlap of NICMOS/HST images, there are multiple measurements for a fraction of stars. In cases of multiple measurements, we computed the mean proper motion for each star, with the corresponding uncertainty given by the error of the mean of the multiple measurements. A comparison between the  uncertainties estimated in this way and the proper motion uncertainties computed as described in the previous paragraph showed that both methods provide similar uncertainties.


\section{Results}
\label{section:Proper motions}

\subsection{A proper motion catalogue for the GC}
\label{subsection:Proper motions catalogue}

Since the proper motions were determined for each of the HAWK-I chips independently, we combined the lists of the four chips corresponding to each GNS pointing. There are some stars with multiple measurements, which can happen when they lie near the inner edges of the chips,\ close to the detector gap that is covered in the observations through jittering. For these stars we computed the mean proper motion and its corresponding uncertainty from the multiple measurements. 

Finally, we combined the proper motions of all the pointings of our study to obtain the proper motion catalogue. Again, for all stars with multiple measurements in the overlapping regions, we calculated their mean proper motions and corresponding uncertainties via averaging. To summarise, we determined the proper motion uncertainties from a combination of astrometric and epoch alignment uncertainties. For stars with multiple proper motion measurements (located in overlapping pointings in the HST and/or the GNS), we determined their proper motion uncertainties directly from the multiple measurements. Both methods result in similar uncertainties.

Figure~\,\ref{fig:measured_pos} shows the positions of the stars in our proper motion catalogue. The different size of the NICMOS images and their tiling pattern, the complex procedure of matching NICMOS data to GNS data, and the increased uncertainties near the edges of the images (both because of lower signal-to-noise ratios and because of larger astrometric transformation uncertainties) mean that the density of our measurements is not homogeneous across the studied field. 

 Figure~\ref{fig:dpm} shows the uncertainties of the measured proper motions as a function of stellar magnitude. There is a small fraction of stars fainter than $H\approx17$. These stars are located in regions with very low stellar densities (i.e. on dark clouds). As is clear from the figure, the proper motion uncertainty of the majority of stars is $<1$\,mas\,yr$^{-1}$, which is due to considering the astrometric uncertainty limit for each of the surveys in the proper motion analysis.

The proper motions of the Quintuplet cluster region in this work have higher uncertainties compared to \cite{Shahzamanian:2019} because we did not use the GNS catalogue data in the latter work but specifically optimised point source detection for the Quintuplet cluster. This also suggests that we may still be able to improve  the quality of our proper motions if we optimise the GNS point source detection pipeline (which goes far beyond the scope of the present work).

\begin{figure}
  \includegraphics[width=\columnwidth]{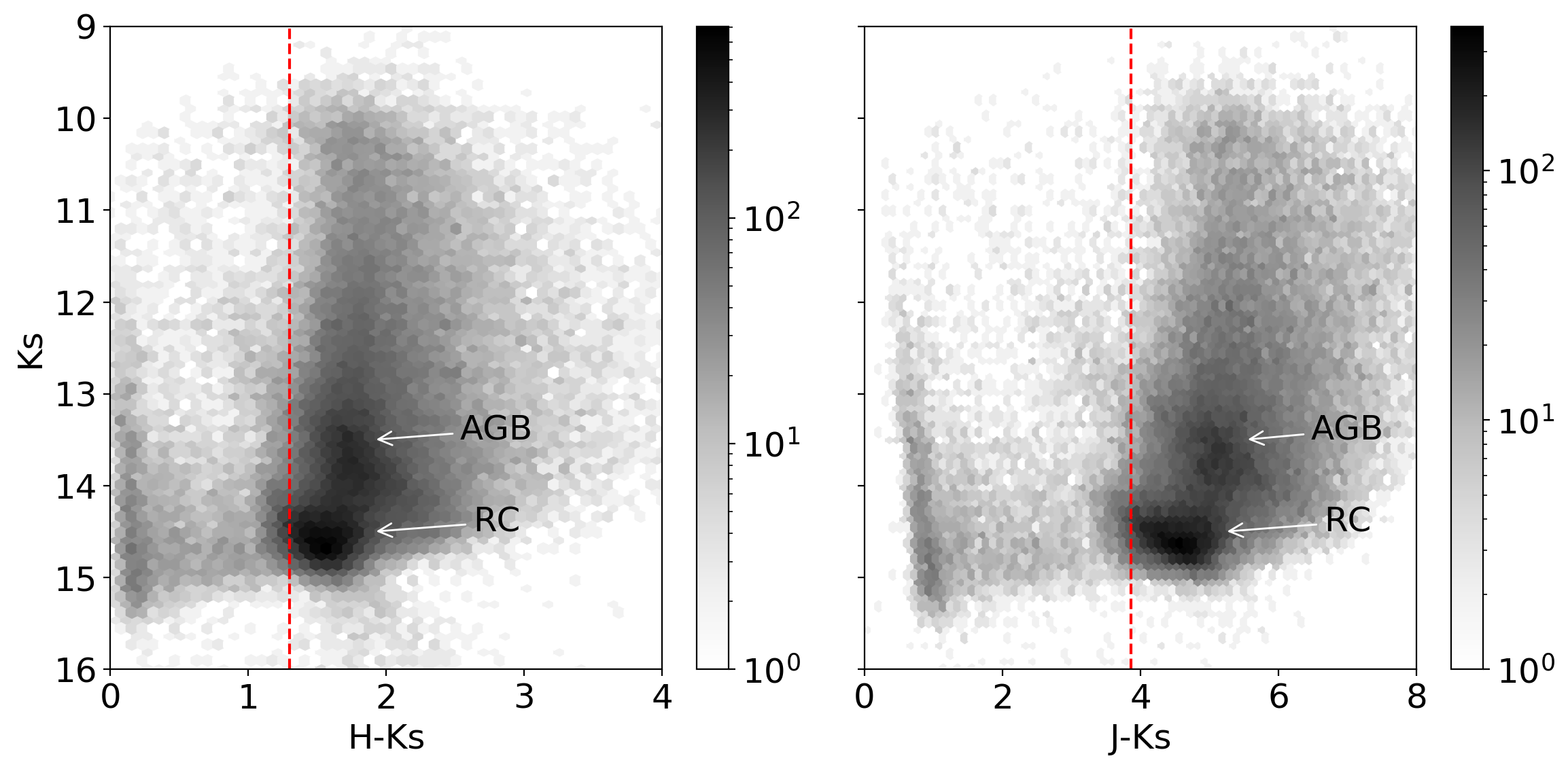}
  \caption{CMDs of the stars in the proper motion catalogue. The dashed red lines illustrate the colour cuts used to remove foreground stars. The arrows mark the asymptotic giant branch (AGB) and the red clump (RC).}
 \label{fig:CMD_hawki}
\end{figure}

Figure~\ref{fig:CMD_hawki} presents CMDs for all stars with measured proper motions using GALACTICNUCLEUS photometry. A colour cut separates foreground stars (mostly Galactic disc and some low-extinction bulge or bar stars) from those in the GC (NSD and inner bulge or bar) via the criteria $H-K_{\rm s} < 1.3$ and $J-K_{\rm s} < 3.8$ \citep[for a justification of the criterion and the extinction curve used, see][]{Nogueras-Lara:2019zv, nogueras:2021}. About 12\% of the stars in our catalogue have $H-K_{\rm s} < 1.3$ and are foreground sources.
The proper motion data only reach down to the top of the red clump. This is a consequence of the relative shallowness of the HST narrow-band images. 

Table~\ref{tab:data_cat} contains the first rows of our proper motion catalogue. We have made the complete catalogue, which contains 77414 objects, available in electronic form via the CDS. The proper motions reported here are relative, not absolute. We established our reference frame for the proper motions by assuming mean zero motion and rotation between the stellar positions extracted from each H-P$\alpha$S pointing and the ones in the corresponding chips and pointings of the GNS.


    \begin{figure}[!b]
      \begin{center}
    
        \subfigure{%
            \includegraphics[width=0.45\textwidth]{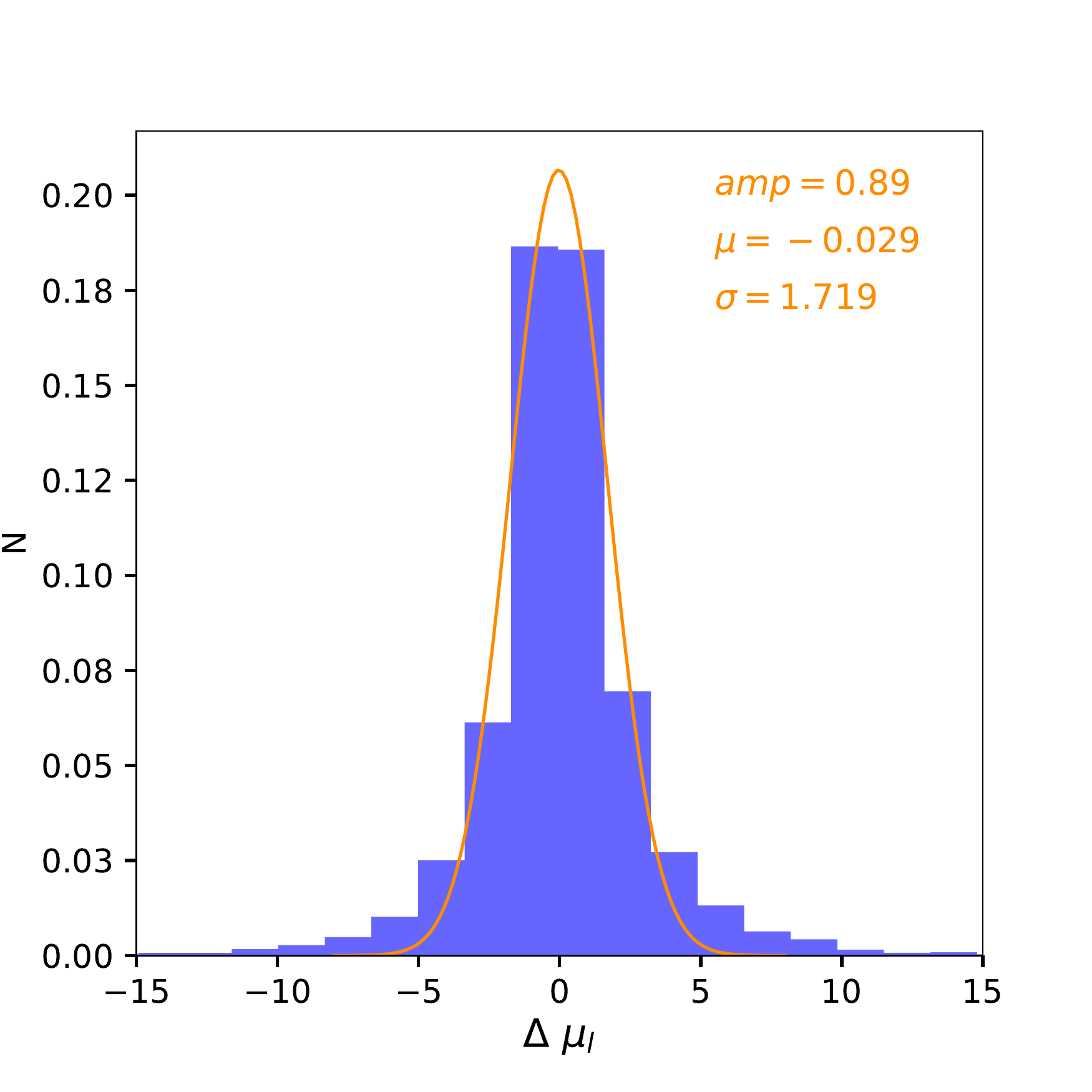}
        }
        \subfigure{%
           \includegraphics[width=0.45\textwidth]{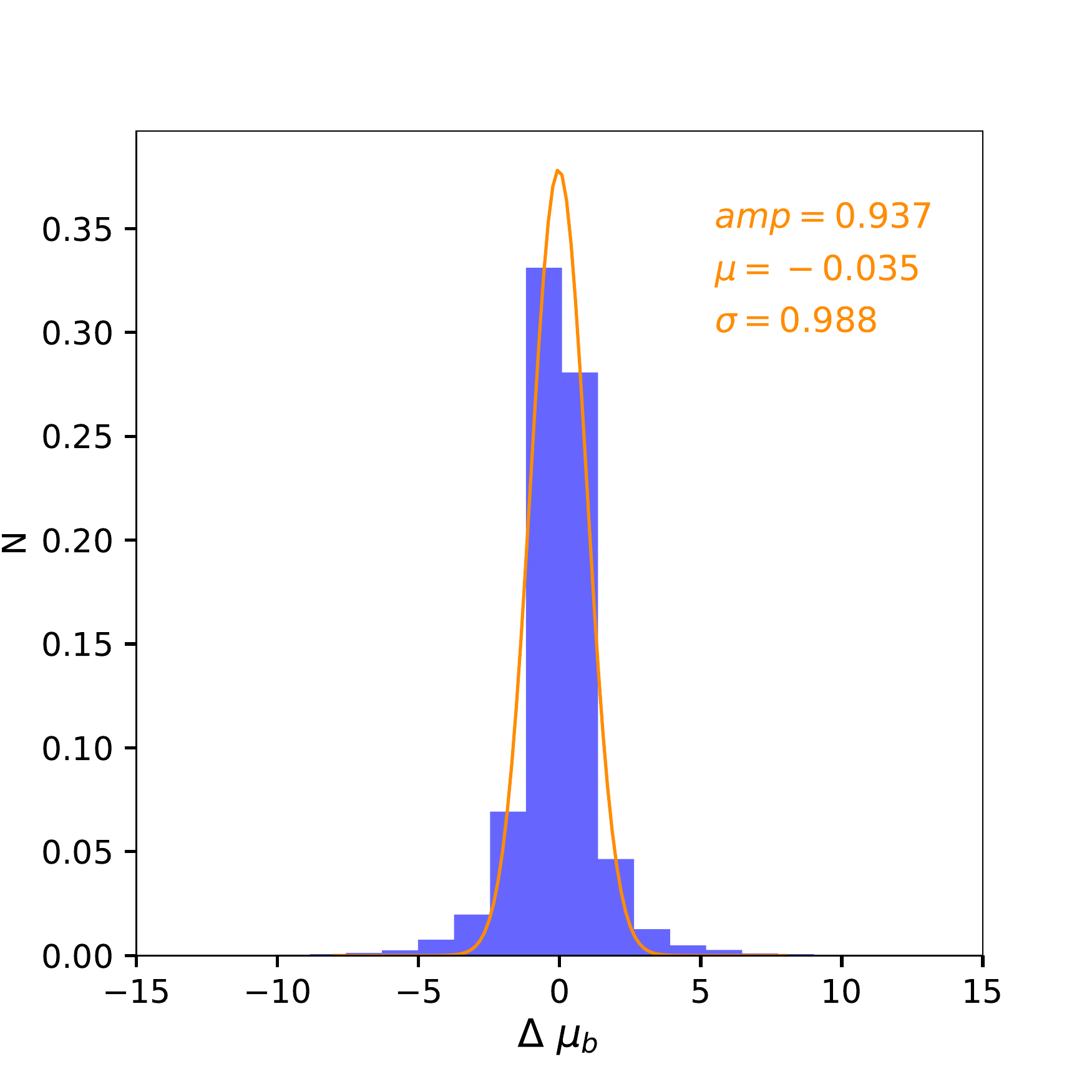}
        }

     \end{center}
    
    \caption{Proper motion differences between our catalogue and the work of \citet{Libralato:2021} measured for each star and normalised by the quadratically combined uncertainties. The orange lines indicate Gaussians fitted to the histograms, and amp, $\mu$, and $\sigma$ are the amplitude, mean, and standard deviation of the Gaussians. 
}
\label{fig:hist_lib}
\end{figure}

\subsection{Verification}

We performed two tests to verify the proper motion measurements of our catalogue.

\subsubsection{Comparison with \citet{Libralato:2021}}
\label{section:Comparison with Libralato work}

\citet{Libralato:2021} present proper motion measurements of stars in the GC acquired with WFC3/HST (F153M) in several non-contiguous areas towards the GC with a total surface area of $\sim144$\,arcmin$^{2}$. Some of their survey area overlaps with ours. We compared the proper motions of stars common to their and our work, using only stars that have an associated proper motion uncertainty in the catalogue of \citet{Libralato:2021}. The two works have consistent photometry, and the cross-identification of sources is straightforward. 

The proper motions given by \citet{Libralato:2021} are absolute (referenced to the Gaia DR2 catalogue), while our proper motions are relative, meaning  that net zero rotation and proper motion for each match between a GNS image and an H-P$\alpha$S image has been assumed. Therefore, in order to compare their proper motions with ours, we obtained the offset between both works and corrected the values for the offset.

The Libralato proper motions are given in equatorial coordinates, while our proper motions are given in Galactic coordinates. We therefore rotated their proper motions to align them with ours. We then computed the differences of the proper motions measured for each star and normalised them by the quadratically combined uncertainties. The histograms of the normalised differences are shown in Fig.~\ref{fig:hist_lib}. We fitted Gaussian distributions to the histograms. They provide very good fits, are centred on a mean difference of zero, and have standard deviations of 1$\sigma$ (b) and 2$\sigma$ (l), respectively. The expected outcome of this comparison under the assumption of purely statistical (random) uncertainties is a Gaussian with mean zero and a standard deviation of 1. 

We conclude that there is good, if not perfect, agreement between the data. A likely cause of systematic uncertainties is that the assumption of zero net motion for our catalogue is probably violated because of the rotation of the NSD combined with differential extinction (see Sect.\,\ref{section:Kinematics in the GC}): We will see more stars on the near side of the NSD than on its far side, which means that our assumption of zero net proper motion is violated to a certain degree. The small FoV of NICMOS may worsen this problem in some fields.

\begin{figure}[!ht]
\includegraphics[width=\columnwidth]{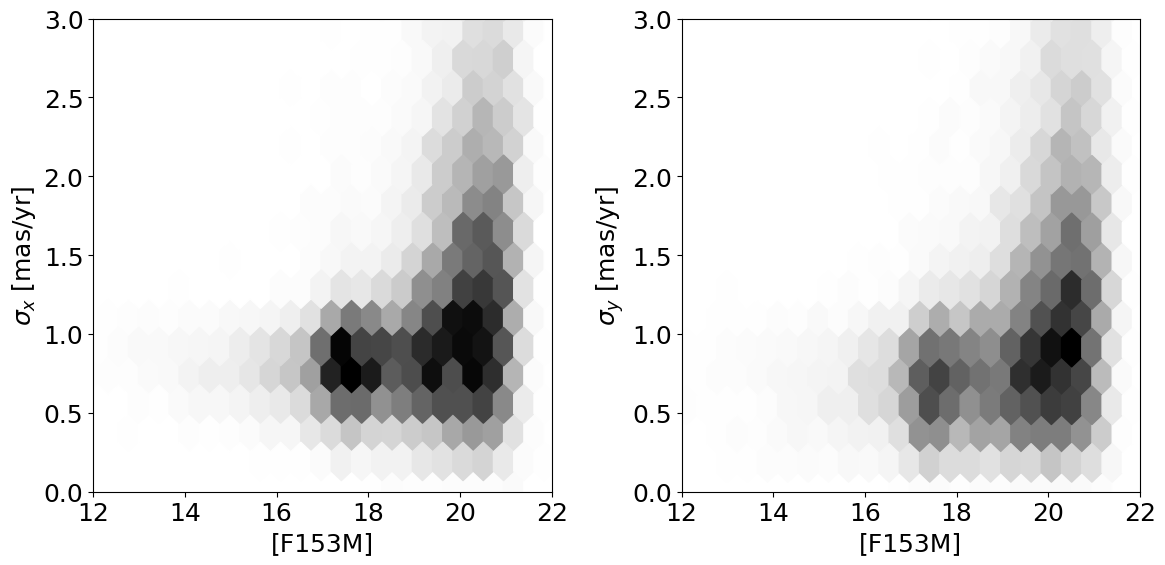}
    \caption{Uncertainties of proper motions measured in HST/WFC3 images of the Quintuplet cluster.}
\label{fig:sigmaQuin}
\end{figure}

\begin{figure}[!ht]
\includegraphics[width=\columnwidth]{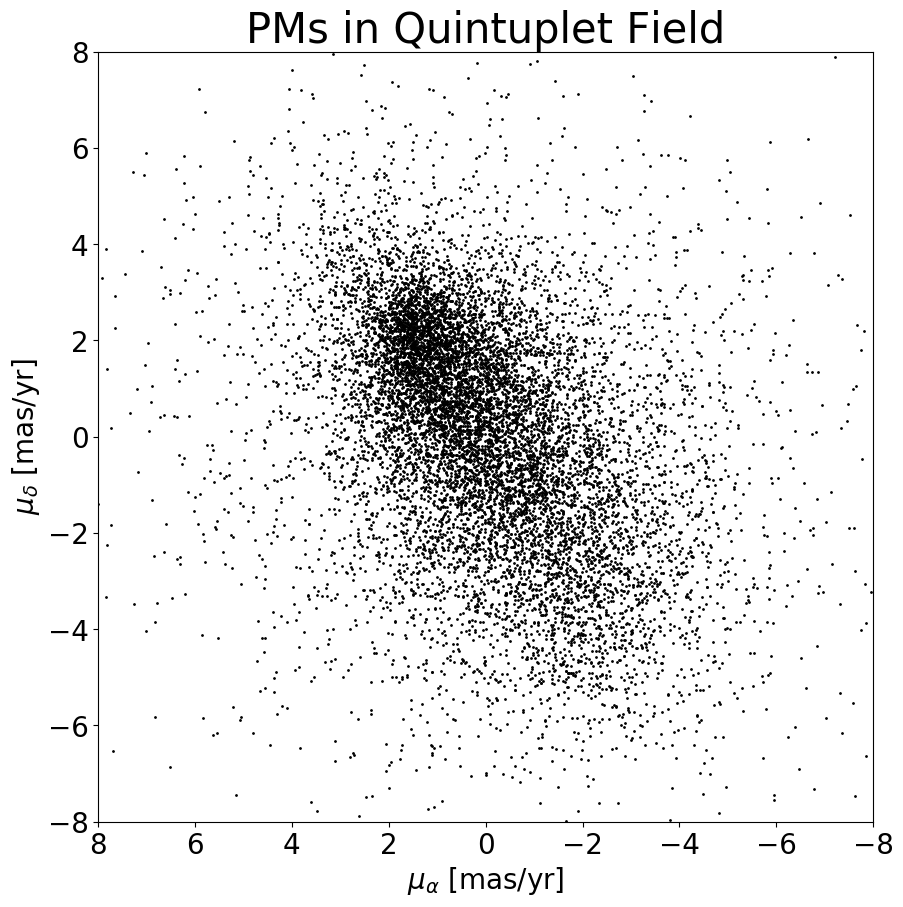}
    \caption{Vector point diagram for proper motions measured in the Quintuplet cluster obtained with HST data.}
\label{fig:VPD_quintuplet}
\end{figure}


\subsubsection{Comparison with HST data for the Quintuplet cluster}
\label{section:Comparison with WFC3/HST data}

Precision proper motions were measured with WFC3/HST imaging on fields centred on the Arches and Quintuplet clusters by \citet{Hosek:2015sh} and \citet{Rui:2019ch}. We downloaded the corresponding HST WFC3 F153M images for the Quintuplet cluster from the Hubble Legacy Archive, which provides fully reduced and astrometrically calibrated images. \citet{Rui:2019ch} provide a description of the observations (see their Sect. 2 and Table\,1). Specifically, we used images from 16 August 2010, 9 September 2011, 12 August 2012, and 22 October 2016. 

We carried out PSF extraction and PSF fitting astrometry and photometry on these images with the \textit{StarFinder} package. We used the 2012 epoch as a reference frame and transformed all stellar positions into this frame via a linear fit (rotation, axis scales, and shear). Common stars were identified by the conditions that they could not lie more than $0.13"$ (one pixel) apart from each other. The alignment procedure and identification of common stars were iterated five times and quickly (after one iteration) converged to a stable set of stars. Finally, proper motions were computed by linear fits to the time-position data. Uncertainties were estimated by assuming uniform weighting and scaling the uncertainties of the fit to a reduced $\chi^{2}$ of one. The uncertainties of the measured proper motions are shown in Fig.~\ref{fig:sigmaQuin}. The precision of our measurements is significantly worse than what has been achieved by \citet{Rui:2019ch} because we have used a far less elaborate procedure, addressing fewer potential sources of systematics. Nevertheless, the measurements are good enough for comparison with our proper motion catalogue. For this purpose, we only used proper motions with an uncertainty of $<1.5$\,mas\,yr$^{-1}$ along each direction.

\begin{figure}[!ht]
\includegraphics[width=\columnwidth]{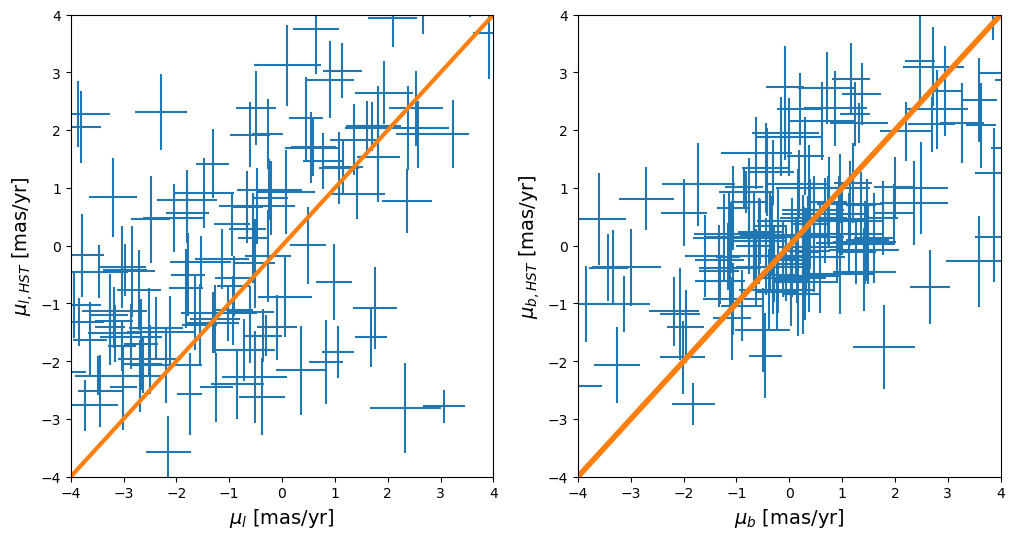}
    \caption{Comparison of proper motions as given in our catalogue ($\mu_{l}, \mu_{b}$) and as measured from WFC3/HST data ($\mu_{l, HST}, \mu_{b, HST}$).}
\label{fig:Quintuplet_comparison}
\end{figure}


We show a vector point diagram of measured proper motions in the Quintuplet cluster in Fig.\,\ref{fig:VPD_quintuplet}. It appears very similar to the one presented in Fig.\,7 (top left) of \citet{Rui:2019ch}. We used a less elaborate method than they did, so the uncertainties and therefore also the scatter of the data points are probably larger. Also, we assumed zero mean motion and rotation for all the stars in the field, while \citet{Rui:2019ch} fixed their frame of reference on the probable Quintuplet cluster members. Therefore, there is an offset between their proper motions and ours.

In Fig.~\ref{fig:Quintuplet_comparison} we compare the proper motions from our catalogue with the proper motions measured on the WFC3/HST data in the Quintuplet field. For the comparison, we used stars that coincided within $0.05"$ in position and corrected for the mean offsets between the two lists ($0.51$\,mas\,yr$^{-1}$ parallel to and $0.24$\,mas\,yr$^{-1}$ perpendicular to the Galactic plane). We only considered stars with proper motion uncertainties of less than $0.8$\,mas\,yr$^{-1}$ in both datasets.

We obtain a Pearson correlation coefficient value, which measures the linear relationship between the two data samples,
of $0.652 \pm 0.016$ for the left plot and $0.624 \pm 0.025$ for the right plot of Fig.~\ref{fig:Quintuplet_comparison}. The uncertainties were obtained using the Monte Carlo method.
The p-values for the correlations in both plots are lower than the significance level of 0.05, which indicates that both correlations are statistically significant. There is still some offset in the left panel of Fig.~\ref{fig:Quintuplet_comparison} that points towards a remaining offset.

We conclude that the agreement is good, considering the uncertainties of the individual measurements.  

    \begin{figure}[!th]
      \begin{center}
    
        \subfigure{%
            \includegraphics[width=\columnwidth]{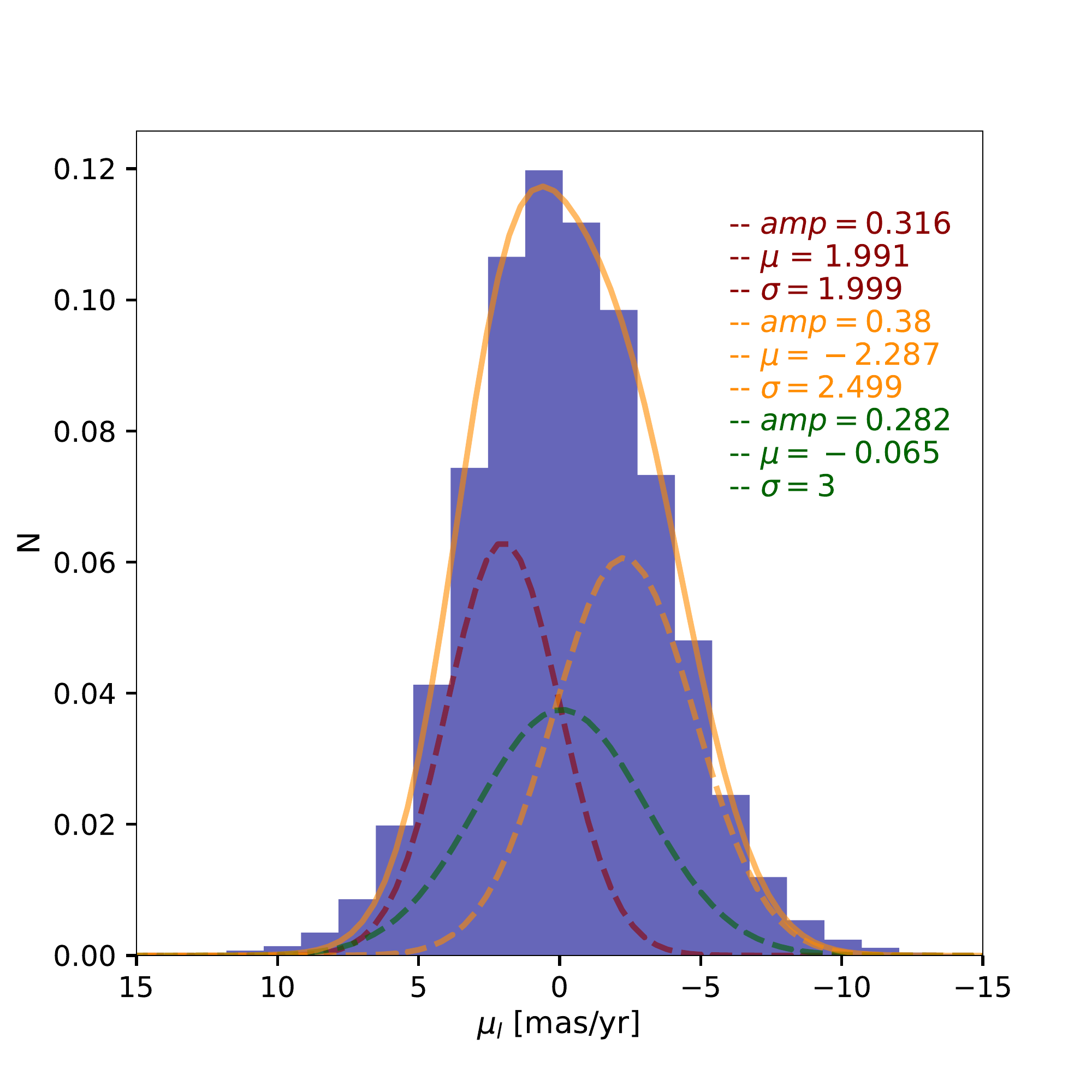}
        }
        \subfigure{%
           \includegraphics[width=\columnwidth]{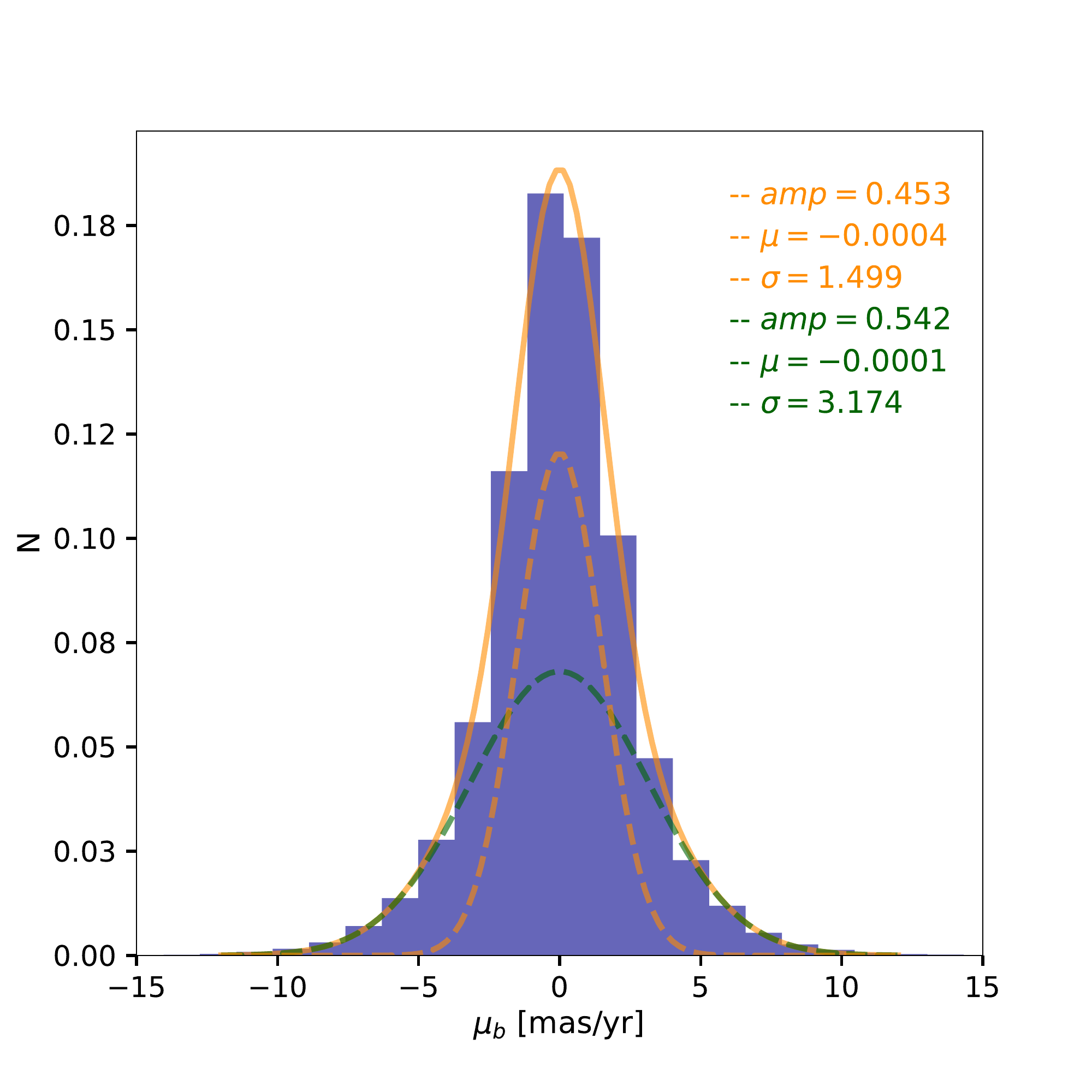}
        }
        \\

     \end{center}
    
    \caption{Normalised velocity distributions of GC stars in the catalogue. The individual Gaussian components are indicated as dashed lines and the global solution as a thick solid line. 
    Top: Direction parallel to the Galactic plane.\  NSD stars moving eastwards are shown in red, NSD stars moving westwards in orange, and bulge stars in green. Bottom: Direction perpendicular to the Galactic plane.\  NSD stars are shown in orange and bulge stars in green.}
\label{fig:hist_cat}
\end{figure}

\section{Observed kinematics }
\label{section:Kinematics in the GC}

\subsection{Kinematics of stars in the GC}
In order to study the kinematics of stars inside the GC, we excluded all foreground stars, which we define as those stars with $H\mbox{-}K_{\rm s}$ < 1.3 \citep[see Fig.~\ref{fig:CMD_hawki} and also][]{nogueras:2019cat}. These stars with low reddening are mostly located in the Galactic disc, but some stars from the near edges of the bulge or bar may also be counted among them.

\begin{table*}[]
\caption{Best-fit values for the velocity distributions of GC stars.}
\centering
\begin{tabular}{c c c c}
\hline
Distribution of $\mu_{l}$ & & & \\
\hline
Component & amp  & $\overline{\mu_{l}}$ & $\sigma_{\mu_{l}}$\\
& &mas\,yr$^{-1}$ & mas\,yr$^{-1}$ \\
\hline
\hline
east & $0.316\pm0.01$ & $1.991\pm0.13$ & $1.999\pm0.01$\\
west & $0.38\pm0.01$ & $-2.287\pm0.15$ & $2.49\pm0.0002$\\
broad & $0.282\pm0.01$ & $-0.065\pm0.06$ & $3\pm0.32$\\
\hline
\hline
Distribution of $\mu_{b}$ & & & \\
\hline
Component & amp  & $\overline{\mu_{b}}$ & $\sigma_{\mu_{b}}$\\
& &mas\,yr$^{-1}$ & mas\,yr$^{-1}$ \\
\hline
narrow & $0.453\pm0.01$ & $-0.0004 \pm 0.0001$ & $1.499\pm0.0002$\\
broad & $0.542\pm0.01$ & $-0.0001\pm0.0001$ & $3.174\pm0.03$\\
\hline
\end{tabular}

\begin{tablenotes}
\small
\textbf{Notes:} amp: amplitudes of different Gaussians; $\overline{\mu_{l}}$: means of Gaussians fitted to the distributions of $\mu_{l}$; $\sigma_{\mu_{l}}$: standard deviations of the Gaussians fitted to the distributions of $\mu_{l}$; $\overline{\mu_{b}}$: means of Gaussians fitted to the distributions of $\mu_{b}$; $\sigma_{\mu_{b}}$: standard deviations of Gaussians fitted to the distributions of $\mu_{b}$; Gaussian function: 
$\frac{amp}{{\sigma \sqrt {2\pi } }}e^{{{ - \left( {x - \mu } \right)^2 } \mathord{\left/ {\vphantom {{ - \left( {x - \mu } \right)^2 } {2\sigma ^2 }}} \right. \kern-\nulldelimiterspace} {2\sigma ^2 }}}.$
\end{tablenotes}

\label{table:data}
\end{table*}

In Fig.~\ref{fig:hist_cat} we show histograms of the proper motions. 
We fitted multiple Gaussians to the observed distributions. In order to constrain the number of Gaussians and to compare different models, we computed the Bayesian model log evidence (ln Z) using the the dynamic nested sampling provided by the \emph{dynesty} package \citep{speagle:2020}. 
We regarded a model to be favoured over another when the difference in their Bayesian log evidence was larger than two \citep{Trotta:2008}. 
The best fitting values and their uncertainties obtained by considering different bin sizes are listed in Table~\ref{table:data}.

A single Gaussian cannot fit the wings of the histogram of $\mu_{b}$ satisfactorily. However, a fit with two Gaussians provides an almost perfect fit (bottom panel of Fig.\,\ref{fig:hist_cat}). Both Gaussians are precisely centred on zero but have significantly different standard deviations, which are measurements of the velocity dispersion. We interpret the broader distribution, with a proper motion dispersion of $\sigma_{\mu_{b,\mathrm{broad}}}=3.17$\,mas\,yr$^{-1}$, as arising from stars from the Galactic bulge because we expect stars from the bulge to be present in our sample and because the velocity dispersion in the bulge was previously determined to have similar values \citep[][]{Clarkson:2008kx,kunder:2012,Soto:2014}. The narrower one, with $\sigma_{\mu_{b,\mathrm{narrow}}}= 1.50$\,mas\,yr$^{-1}$, then corresponds to stars in the NSD, which have a smaller velocity dispersion perpendicular to the Galactic plane than bulge stars. 

\begin{figure*}[!htb]
\includegraphics[width=\textwidth]{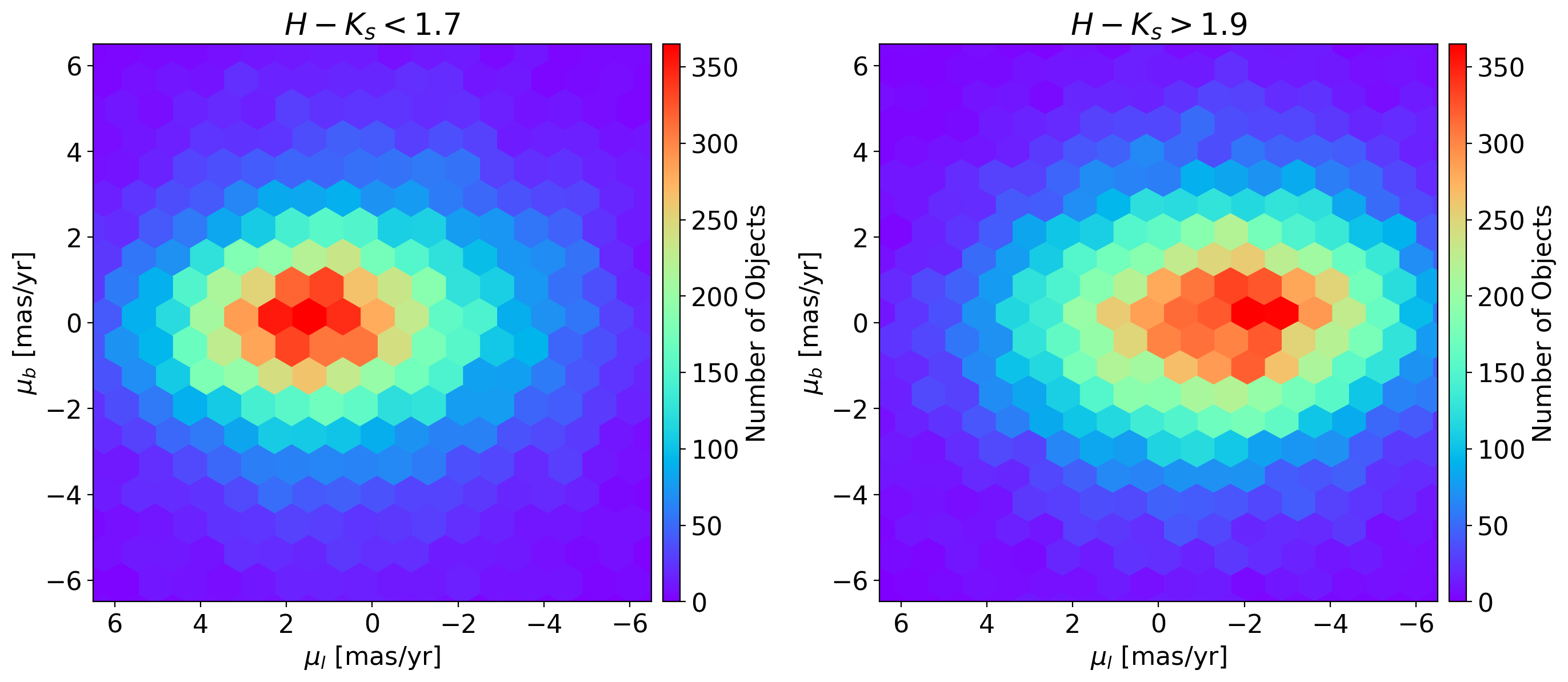}
    \caption{Proper motion density plot for stars with low (left) and high (right) reddening.}
\label{fig:rotation}
\end{figure*}


 This is consistent with previous work, which suggests that we should expect to distinguish two populations of stars by their velocity dispersion perpendicular to the Galactic plane: a broad distribution for bulge stars and a narrower one for NSD stars \citep{Schoenrich:2015}. The different velocity dispersions of the metal-rich and metal-poor stars in \cite{schultheis2021} show the mixture and co-penetration of kinematically cooler NSD stars and hotter stars from the bulge. High precision proper motion studies with WFC3/HST on fields containing the Arches and the Quintuplet clusters indicate that there are indeed stellar populations that may be identified from their kinematics as pertaining to the NSD and bulge or bar, with $\sigma_{b,\mathrm{NSD}}=0.6-0.8$\,mas\,yr$^{-1}$ and $\sigma_{b,\mathrm{Bulge}}=3.0-3.4$\,mas\,yr$^{-1}$, respectively \citep[Field Gaussians 1 and 2-3 in][]{Hosek:2015sh,Rui:2019ch}. These studies also indicate the potential existence of an additional population with intermediate properties.

The rotation of the nuclear disc broadens the velocity distribution along Galactic longitude, $\mu_{l}$. Although the histogram of the velocities parallel to the Galactic plane can be fit satisfactorily with two Gaussians, we require a third distribution to be present, corresponding to bulge stars, which must also be present in this histogram. The superposition of three Gaussians provides a perfect fit to the data (top panel of Fig.\,\ref{fig:hist_cat}). The broad Gaussian centred on a mean motion of $\overline{\mu_{l,\mathrm{broad}}}\approx0$\,mas\,yr$^{-1}$ and with a dispersion of $\sigma_{\mu_{l,\mathrm{broad}}}\approx3$\,mas\,yr$^{-1}$ represents the bulge stars.


    \begin{figure*}[]
      \begin{center}
    
        \subfigure{%
            \includegraphics[width=0.32\textwidth]{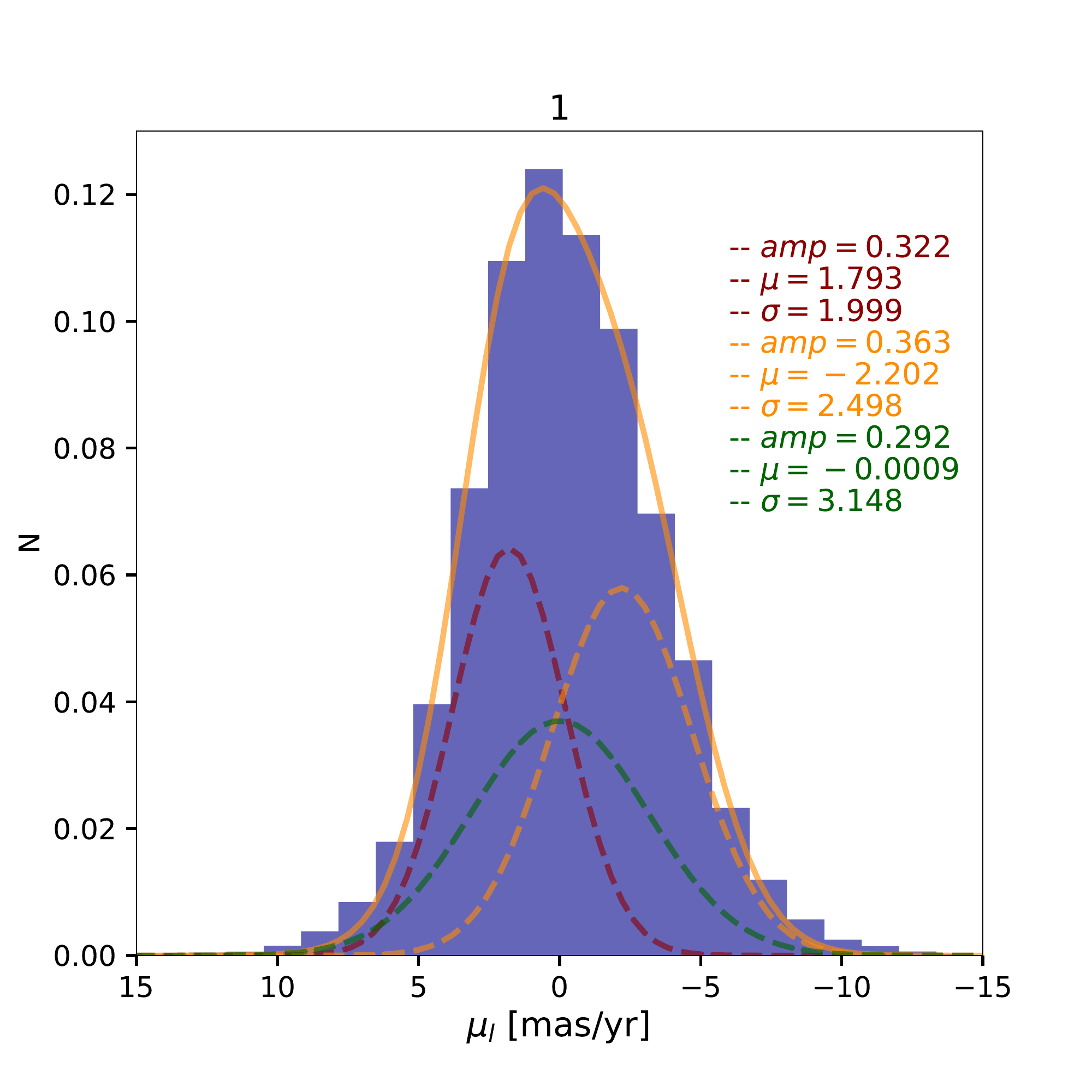}
        }
        \subfigure{%
           \includegraphics[width=0.32\textwidth]{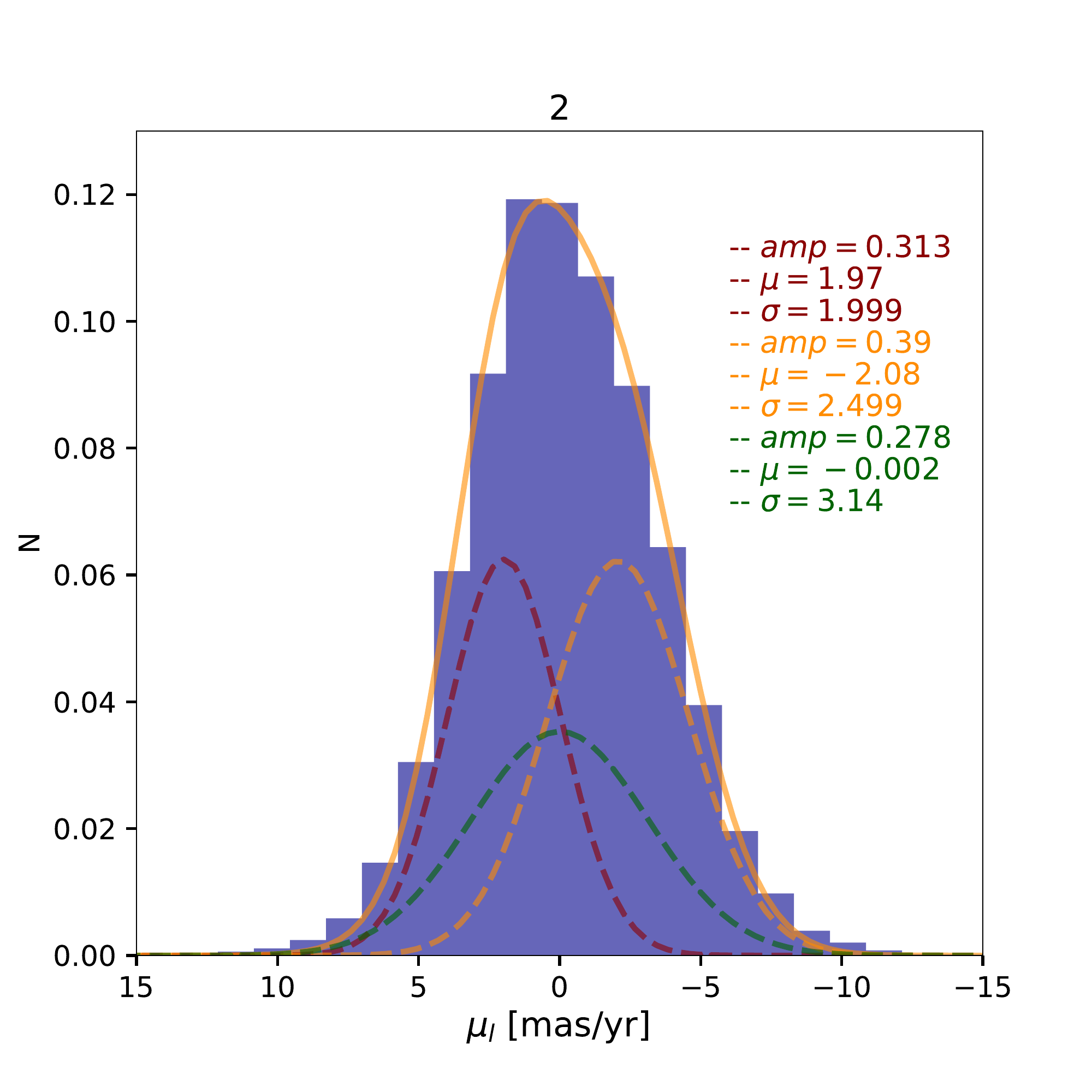}
        }
        \subfigure{%
           \includegraphics[width=0.32\textwidth]{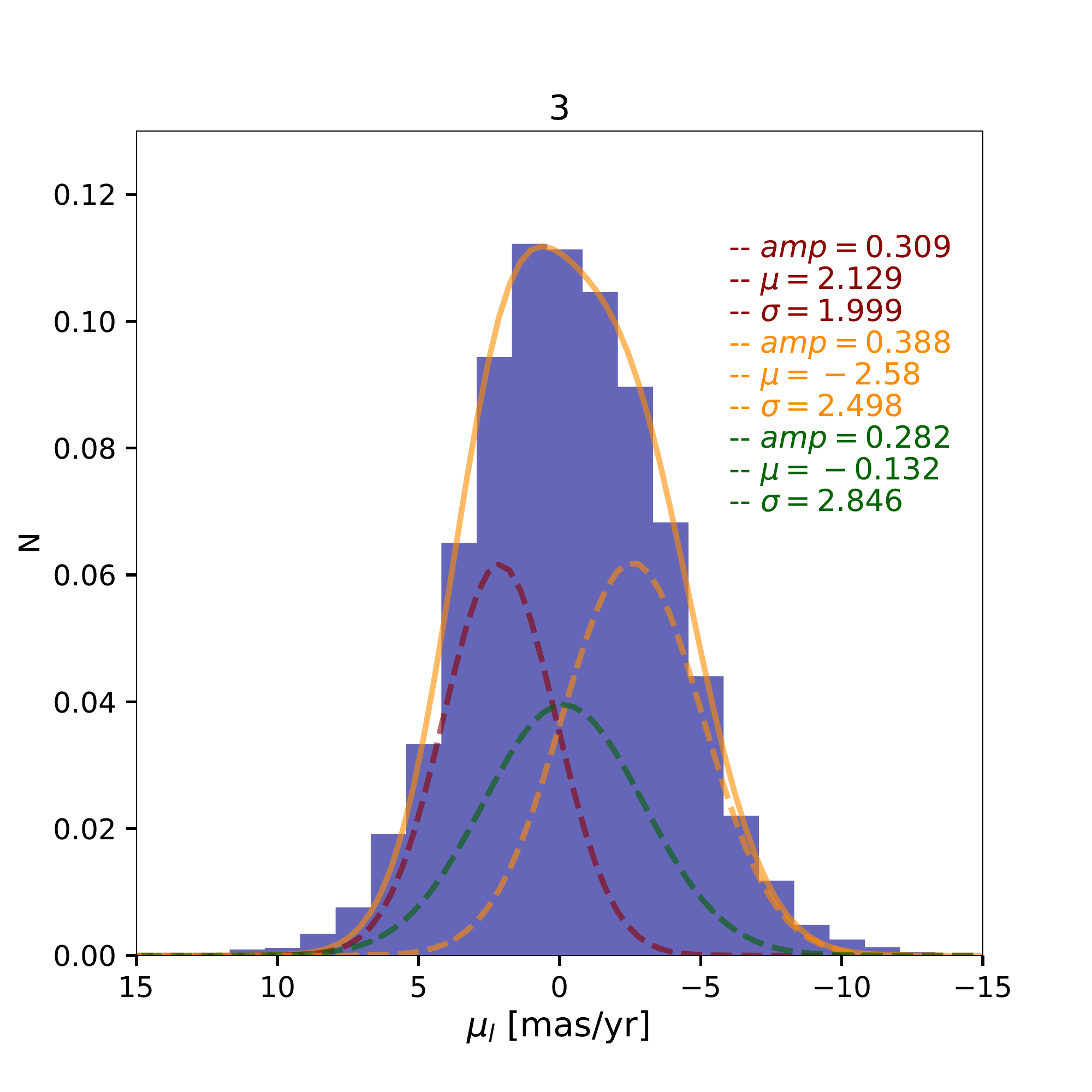}
        }
        \\
        \subfigure{%
            \includegraphics[width=0.32\textwidth]{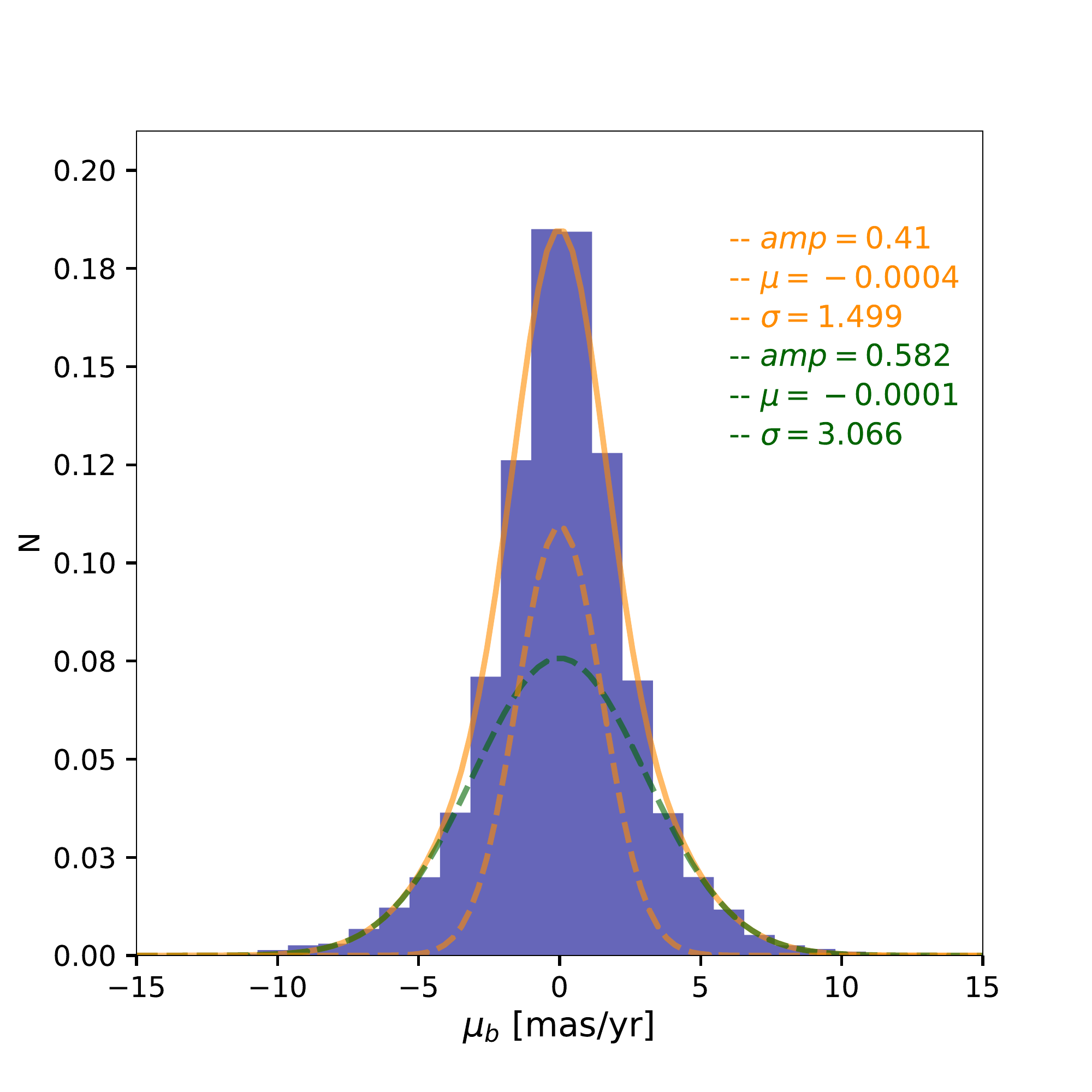}   
        }
        \subfigure{%
            \includegraphics[width=0.32\textwidth]{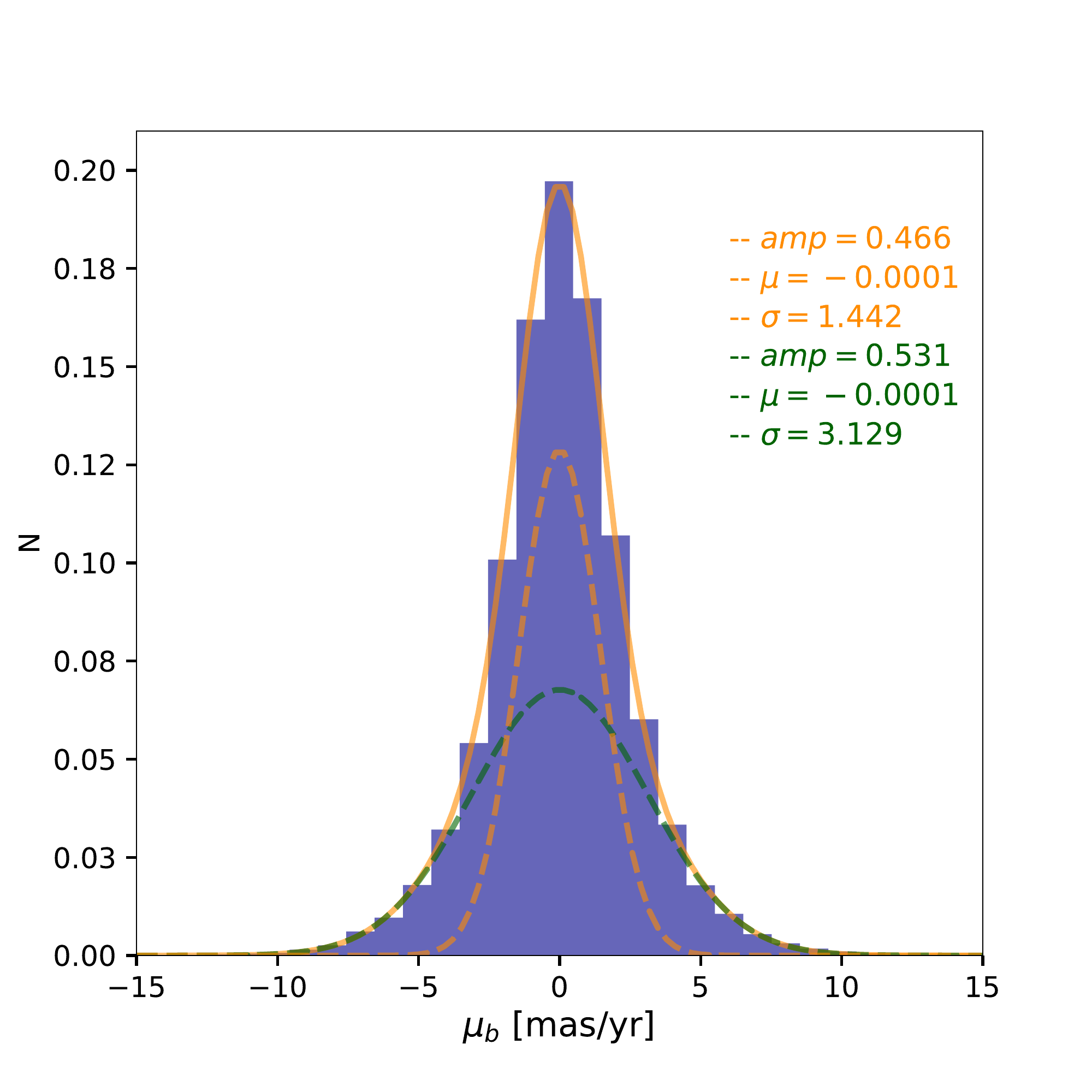}
        }
        \subfigure{%
            \includegraphics[width=0.32\textwidth]{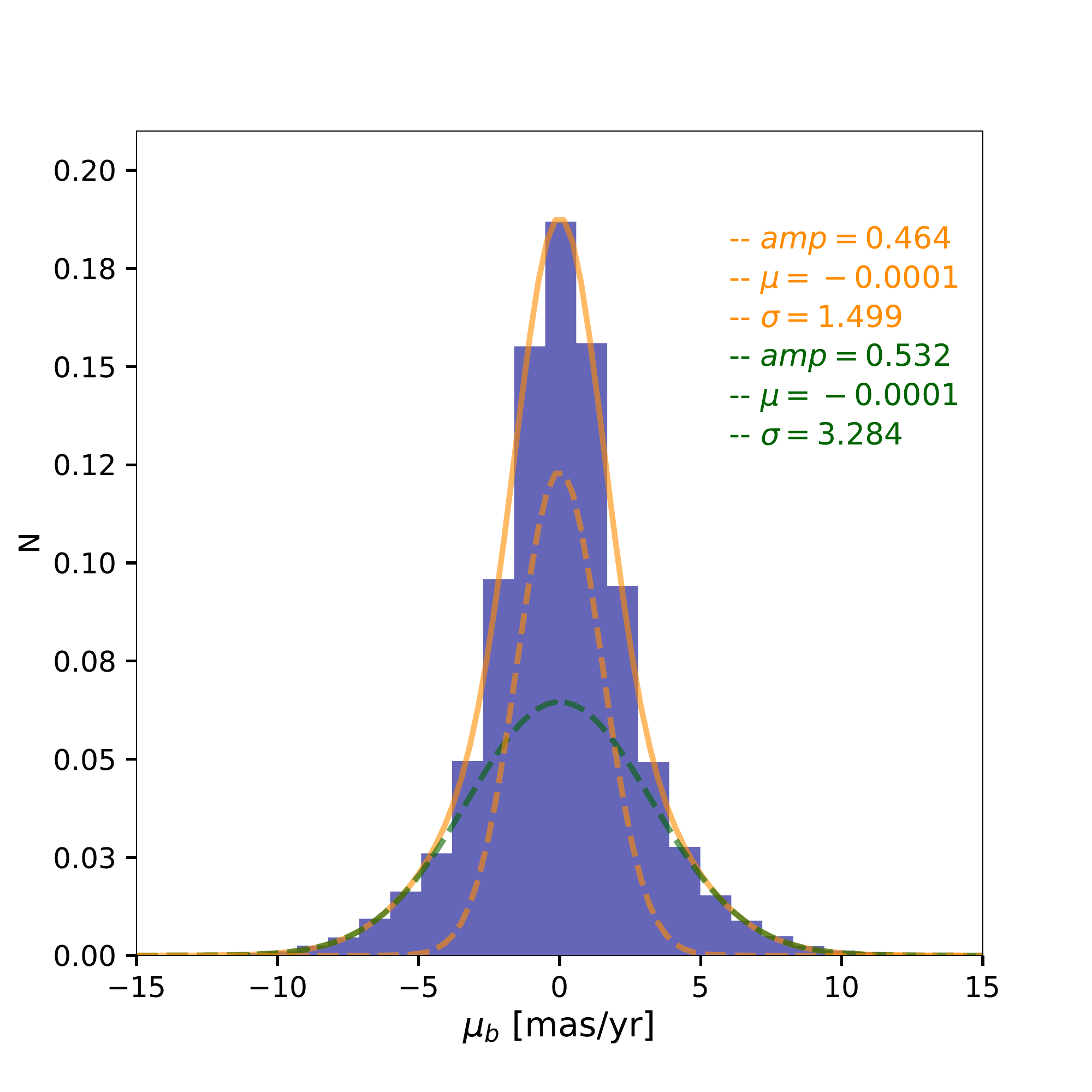}
        }
        \\

     \end{center}
    
    \caption{Normalised velocity distributions of GC stars for three sub-regions of the catalogue from the vertical division. The first, second, and third columns correspond to sub-regions 1, 2, and 3, respectively, progressing from south to north. The individual Gaussian components are presented as dashed lines and the global solution by a thick solid line. Upper panels: Direction parallel to the Galactic plane.\ NSD stars moving eastwards are shown in red, NSD stars moving westwards in orange, and bulge stars in green. Lower panels: Direction perpendicular to the Galactic plane. NSD stars are shown in orange and bulge stars in green. 
        }
\label{fig:hist_cat_vertic}
\end{figure*}


We interpret the other two Gaussians, with mean proper motions of $\overline{\mu_{l,\mathrm{east}}}=1.99$\,mas\,yr$^{-1}$ and 
$\overline{\mu_{l,\mathrm{west}}}=-2.28$\,mas\,yr$^{-1}$ and velocity dispersions $\sigma_{\mu_{l,\mathrm{east}}}=1.99$\,mas\,yr$^{-1}$ and $\sigma_{\mu_{l,\mathrm{west}}}=2.49$\,mas\,yr$^{-1}$, as stars in the rotating NSD. Their mean velocities are almost equal within uncertainties in opposite directions, and therefore the NSD is symmetric. The Gaussian with a broader distribution corresponds to stars moving eastwards (positive Galactic longitude) and the other one to stars moving  westwards. Stars moving eastwards are located preferentially on the near side of the NSD and those moving westwards at its far side. This corresponds to the direction of Galactic rotation and the rotation of the NSC and of the NSD \citep{Trippe:2008it,Schodel:2009zr,Feldmeier:2014kx,Chatzopoulos:2015uq,Lindqvist:1992fk,Schoenrich:2015,schultheis2021}. 

The mean values of eastward and westward proper motions correspond to 80\,km\,s$^{-1}$, in good agreement with the rotation velocity of the NSD that has been derived via near-infrared spectroscopic observations of giants \citep{Schoenrich:2015} and radio observations of OH/IR stars \citep{Lindqvist:1992fk}.
Due to differential extinction along the line-of-sight through the NSD, we expected to detect more stars at the near side of the NSD that have a higher amplitude of the Gaussian that corresponds to the population with positive (eastward) $\overline{\mu_{l}}$. Although we cannot say that we detect more eastward than westward moving stars, it is still true that the mean movement depends on the reddening selection. We can test this prediction by producing proper motion density plots for GC stars with low and high reddening. We made cuts in colour to have two groups, low ($1.3<H-K_{s}<1.7$) and high ($1.9<H-K_{s}$) reddening stars. These plots are shown in Fig.\,\ref{fig:rotation} and confirm that stars with lower extinction move preferentially eastwards, while those with higher extinction move preferentially westwards.

To study whether the velocity distributions change as a function of Galactic latitude, we divided our catalogue into three equal areas parallel to the Galactic plane, one on the plane and the other two above and below it (see Fig.~\ref{fig:measured_pos}). Again, we fitted Gaussians to the distributions with the {\it dynesty} package as described above. The distributions with their fits are shown in Fig.~\ref{fig:hist_cat_vertic}.  In all cases, the $\mu_{l}$ distribution can be fitted with three Gaussians and the $\mu_{b}$ distribution with two. The velocity dispersions of the nuclear disc and bulge populations do not change in any significant way. 
Also, the mean velocities of the eastward and westward moving populations in the nuclear disc remain equal within uncertainties in different regions.

\subsection{Kinematics of foreground stars}

\begin{figure*}[]
      \begin{center}
    
        \subfigure{%
            \includegraphics[width=0.45\textwidth]{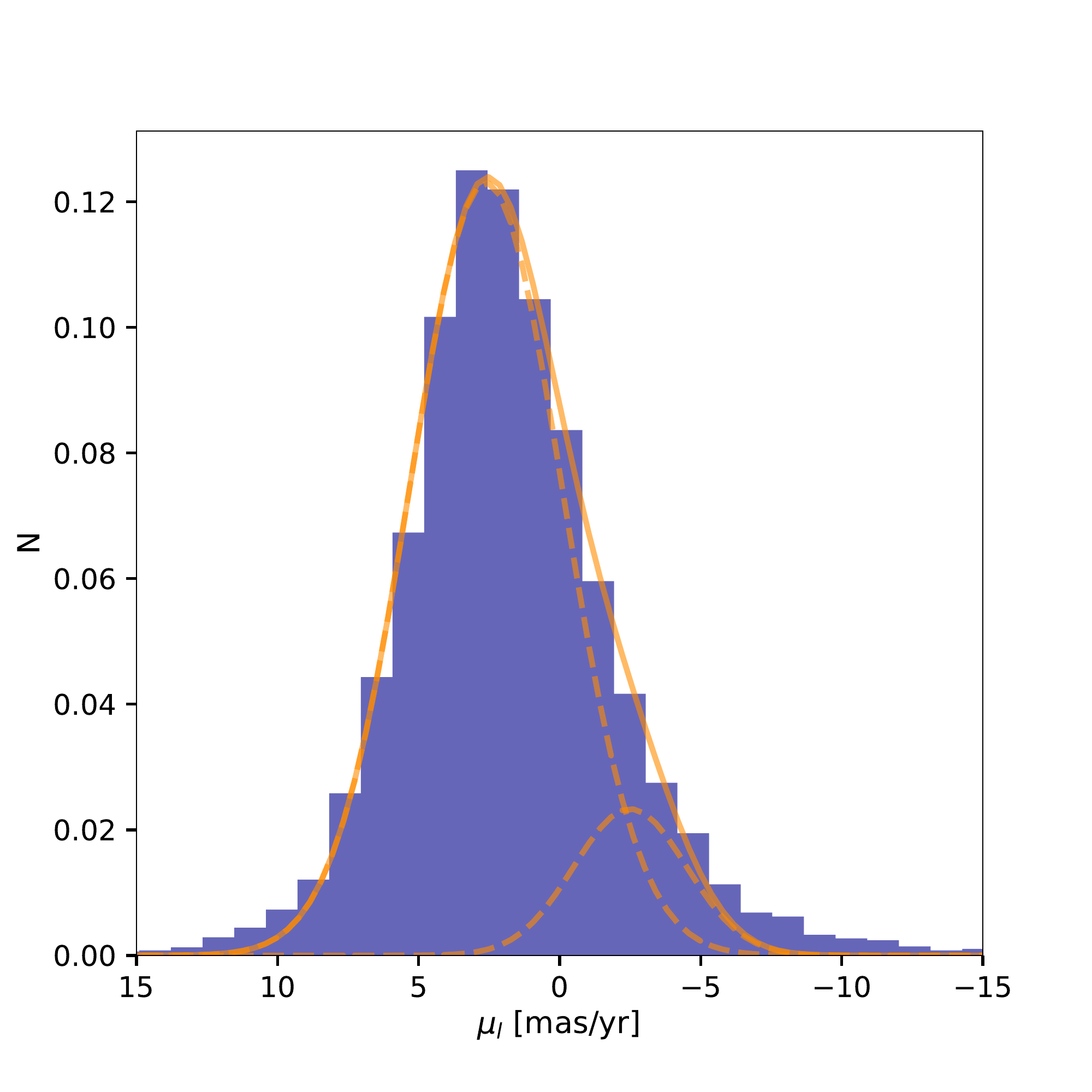}
        }
        \subfigure{%
           \includegraphics[width=0.45\textwidth]{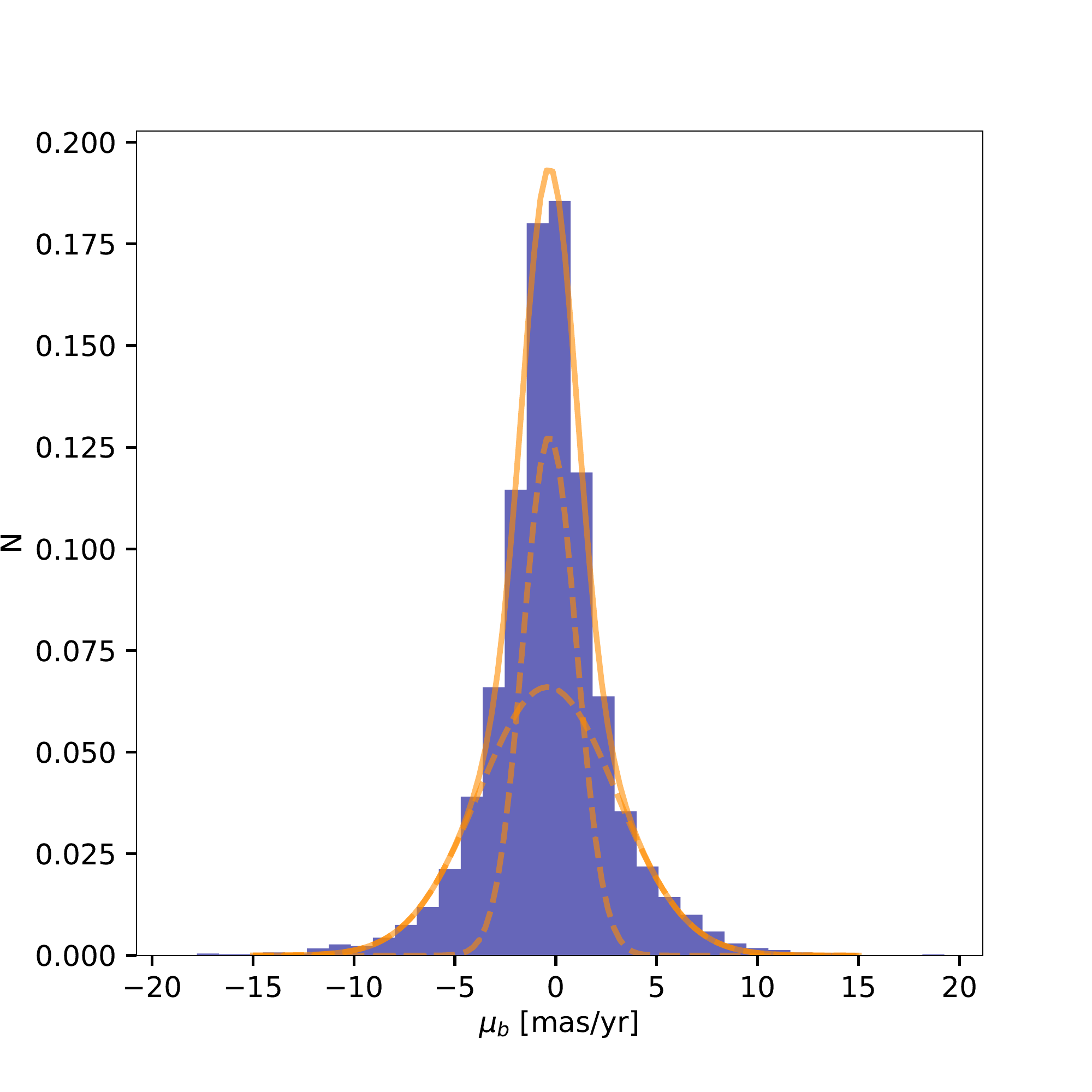}
        }
        \\

     \end{center}
    
    \caption{Normalised velocity distributions of foreground stars.
    Left: In the direction parallel to the Galactic plane. Right: In the direction perpendicular to the Galactic plane.}
\label{fig:hist_cat_foreground}
\end{figure*}

Figure~\ref{fig:hist_cat_foreground} shows the velocity distribution of the foreground stars ($H-K_{s}<1.3$). The velocity distributions of these stars are different from the ones for the GC stars. In particular, there is a significant mean motion towards the east. This corresponds to the direction of Galactic rotation and therefore with aligns with our expectation for the kinematics of stars in the Galactic disc. The $\mu_{l}$ distribution is asymmetric and can be fitted with two Gaussians. A single Gaussian does not appear to be sufficient to fit the $\mu_{b}$ distribution satisfactorily. The interpretation of the foreground population is complex, among other reasons because the foreground stars are placed at significantly different distances from Earth. Since in this work we are concerned with the stellar populations in the GC, we will not further discuss the kinematics of the foreground population.

\section{Finding co-moving groups}
\label{section:Finding co-moving groups}

We applied the unsupervised clustering algorithm `density-based
spatial clustering of applications with noise' \citep[DBSCAN;][]{Ester96} to search for co-moving groups of stars in our data, using the Python implementation\footnote{http://scikit-learn.org/stable/modules/generated/sklearn.\\cluster.DBSCAN.html}. This method has been successfully applied to Gaia DR2 data to find new open clusters \citep[e.g.][]{Castro-Ginard:2018, beccari:2018}. DBSCAN can handle noise because not all the points are assigned to a cluster. It can also detect arbitrarily shaped clusters and does not need any prior information about the expected number of substructures in the data. Here, we show the application of this algorithm to the data from chip 2 of GNS pointing 10, described in \cite{Shahzamanian:2019}, because the Quintuplet cluster is located in this region.

DBSCAN is based on two parameters: a minimum number of points ($N_{min}$) and a length scale ($\epsilon$). A hypersphere with a radius of $\epsilon$ is centred on each star, and the points are regarded to be clustered if the number of stars that are inside the hypersphere is equal to or greater than the pre-determined $N_{min}$. There are three types of points: core, border, and outlier. A point is a core point if at least $N_{min}$ number of points are within radius $\epsilon$, while a border point is a point that is reachable from a core point and has fewer than $N_{min}$ number of points within its surrounding area. If a point is not a core point and cannot be reached from any core points, it is labelled as an outlier. A cluster includes core points that are reachable from one another and all of their border points.

Here, in order to reduce the free input parameters of the algorithm, we obtained the $\epsilon$ value by using the Nearest Neighbors (KDTree algorithm) implementation on scikit-learn \citep{pedregosa2011}. A minimum threshold of ten neighbours per source was considered, and the distances to the neighbours of each source is returned by the algorithm. The optimal epsilon value is the knee value of the nearest neighbours distances plot, determined from the \textit{kneed} python package\footnote{https://pypi.org/project/kneed/}. 

\begin{figure*}[]
   \centering
  \subfigure{%
         \includegraphics[width=0.48\textwidth]{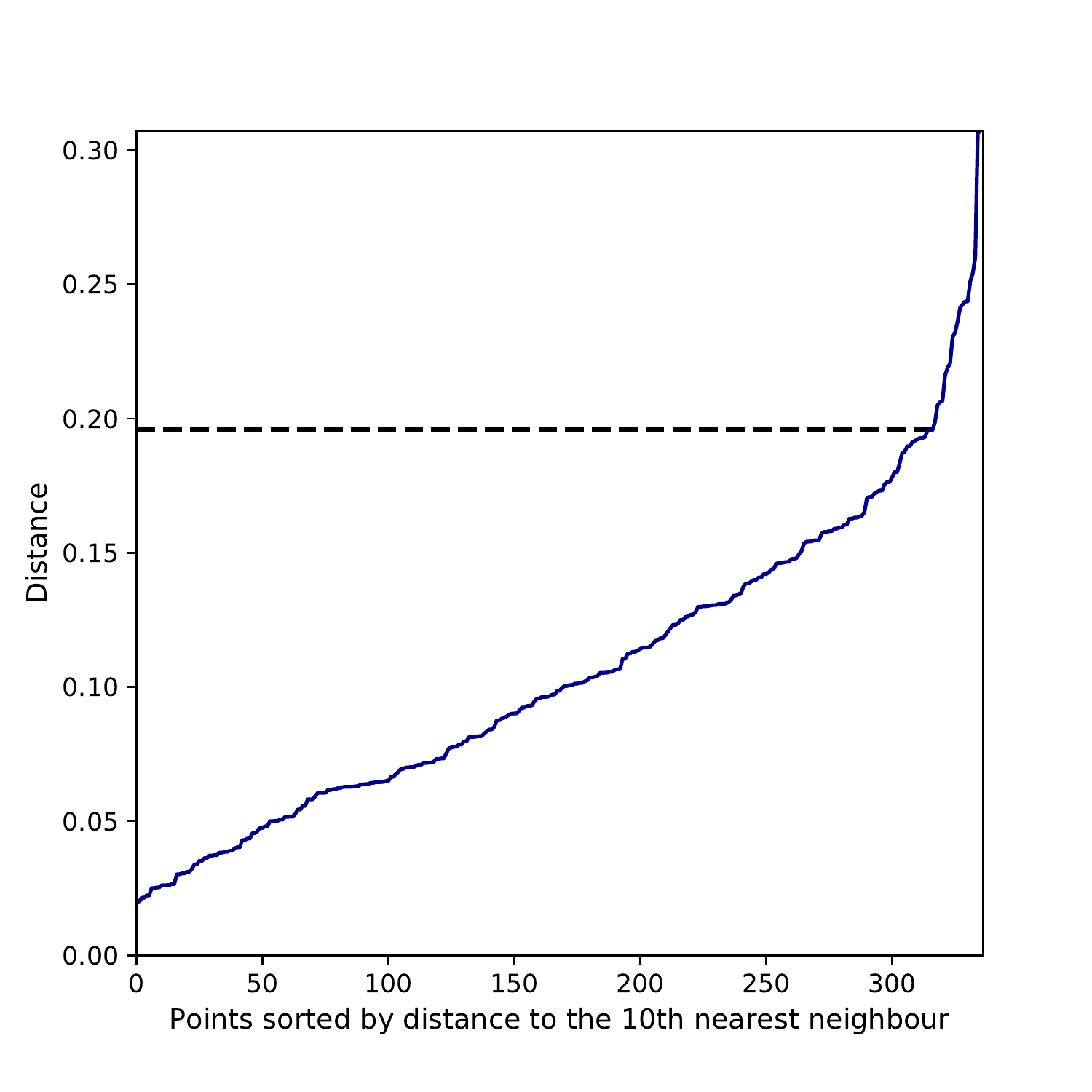}
         }
         \subfigure{%
           \includegraphics[width=0.48\textwidth]{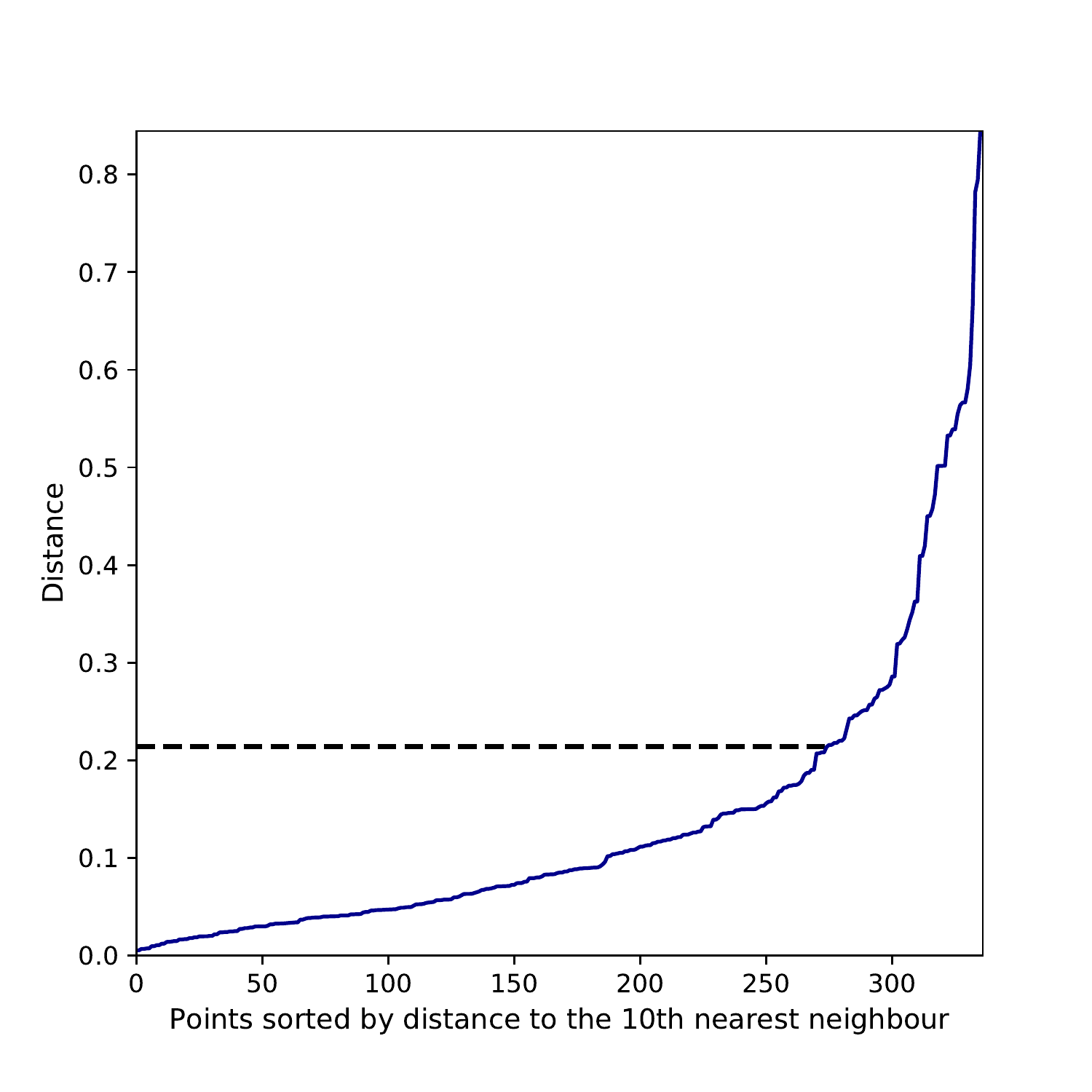}
        }\\
  \caption{Points sorted by distance to the tenth nearest neighbour in position space (left) and velocity space (right). The distance is based on the standardised data. The dashed line shows the knee value of the plot from which the best length scale is determined.
 }

  \label{fig:epsilon} 
\end{figure*}

We used stars with the astrometric uncertainty cut that we applied in the beginning of our proper motion analysis. We first ran the DBSCAN algorithm on the position space of the image (x, y), after standardising the data so that they all have a mean of zero and a standard deviation of 1, using an $\epsilon$ of $\sim$0.2, obtained from running the KDTree algorithm (see left panel of Fig.~\ref{fig:epsilon}), and an $N_{min}$ of 10. We distinguished two groups in the position space (see the top panel of the left column of Fig.~\ref{fig:dbscan_plots}): the more populated one is the Quintuplet cluster, and the one with the lower density is a potential newly found group. In the bottom panel of the left column of Fig.~\ref{fig:dbscan_plots}, the proper motions of the stars belonging to the two groups found in the position space are shown. In this panel, one can see that the Quintuplet cluster sources show a relatively large scatter, and in spite of this, most of the stars of the new group fall outside the area covered by them. Also, only a few stars of the new group share the same kinematics as the Quintuplet cluster.

%


We also checked clustering in the velocity space ($\mu_{l}$, $\mu_{b}$) by running the DBSCAN algorithm using an $\epsilon$ of $\sim$0.2 (see the right panel of Fig.~\ref{fig:epsilon}) and an $N_{min}$ of 10. Here we identify the Quintuplet cluster; however, we cannot detect the potential new group in this space (see the top panel of the right column of Fig.~\ref{fig:dbscan_plots}). The positions of the stars selected in the velocity space are shown in the lower plot of the right column of Fig.~\ref{fig:dbscan_plots}, which includes not only the positions of Quintuplet cluster sources but also some other sources that happen to have proper motions close to these cluster sources. Since the new group is detected in the position space and not the velocity space, we needed to check these spaces separately and not search for clusters in 4D space (positions and velocities) in our data. 

The reason that we cannot detect the new group of stars in the velocity space might be due to its movement parallel to the Quintuplet cluster. The new group is probably part of the Quintuplet cluster since it comprises few stars and lies not very far from the cluster centre. The projected distance of this group from the Quintuplet cluster centre is about 1.3 times larger than the half-light radius of the cluster. 

\begin{figure}[!ht]
\includegraphics[width=\columnwidth]{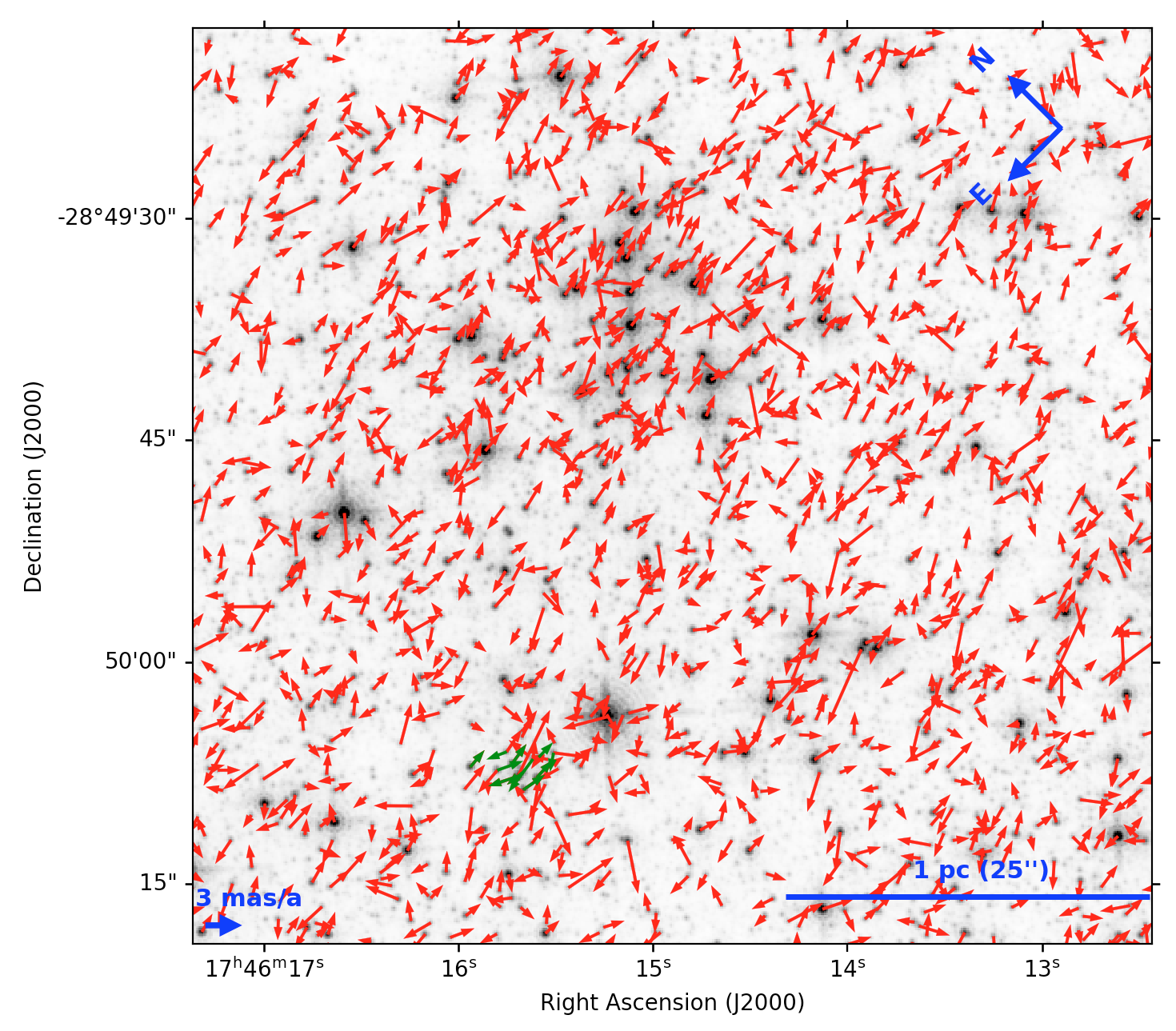}
    \caption{Proper motion measurements of the Quintuplet cluster region. The new co-moving group of stars is shown with green arrows.}
\label{fig:quin-hst-overplot}
\end{figure}

\begin{figure*}[]
     \begin{center}
    
        \subfigure{%
            \includegraphics[width=0.48\textwidth]{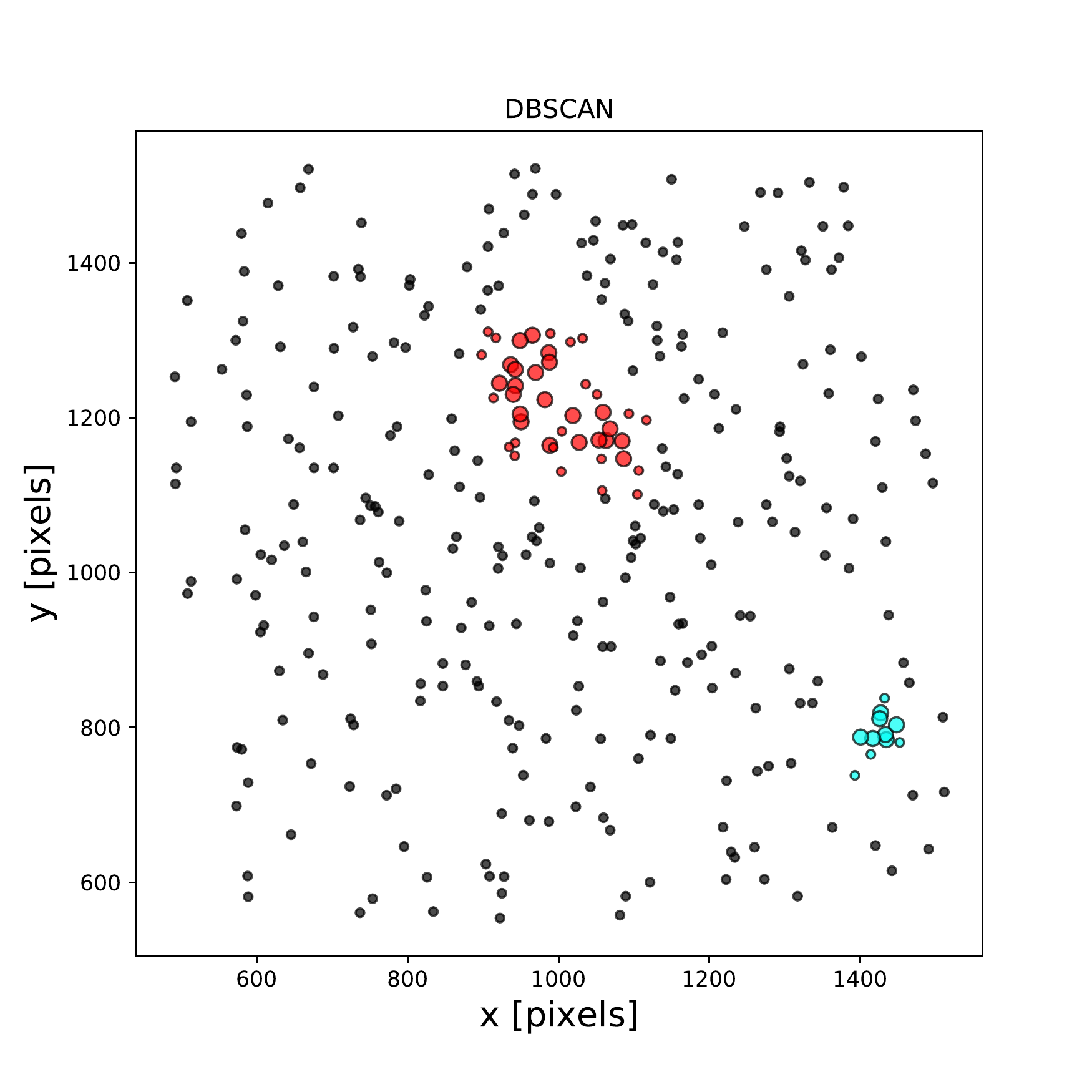}
        }
         \subfigure{%
           \includegraphics[width=0.48\textwidth]{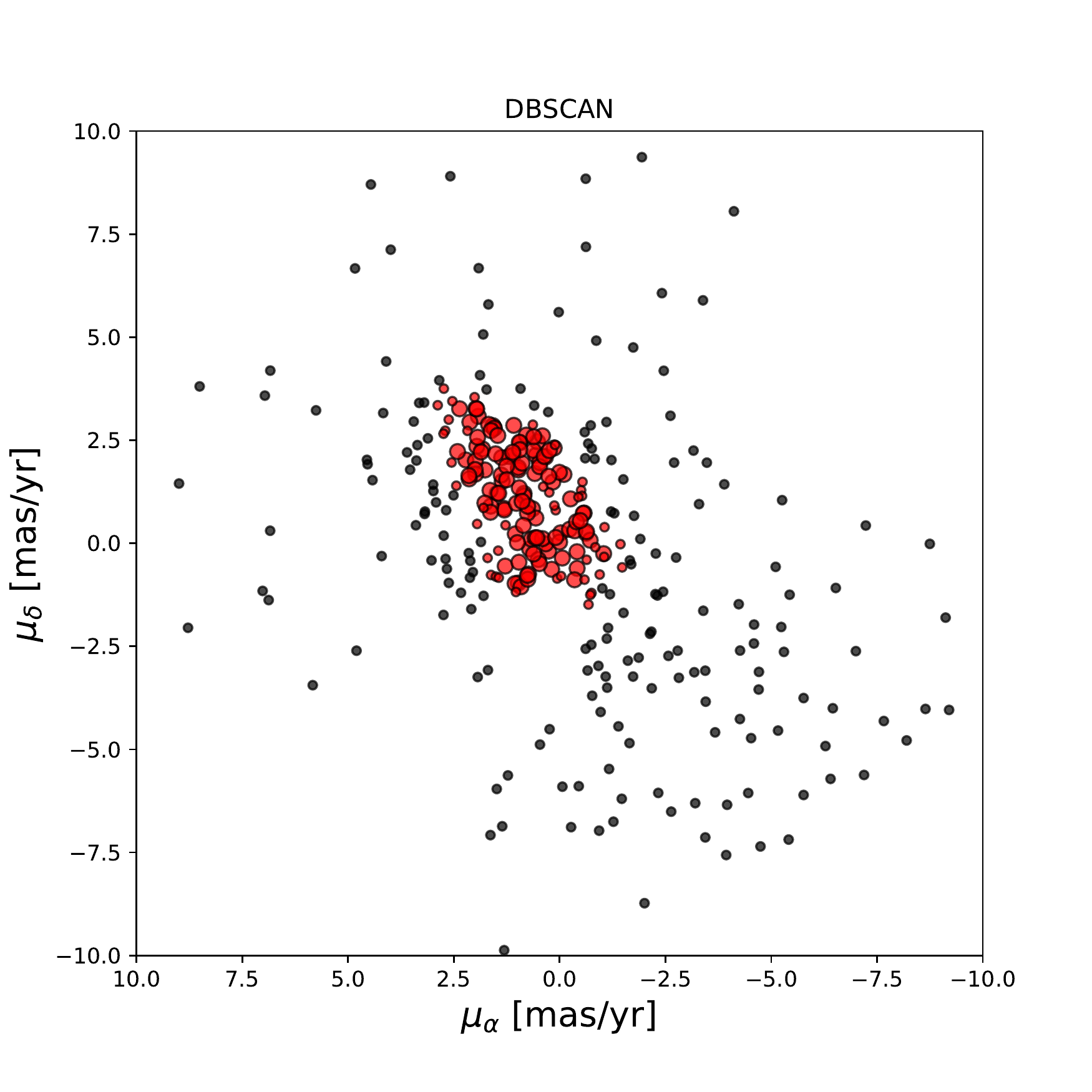}
        }\\
        \subfigure{%
           \includegraphics[width=0.48\textwidth]{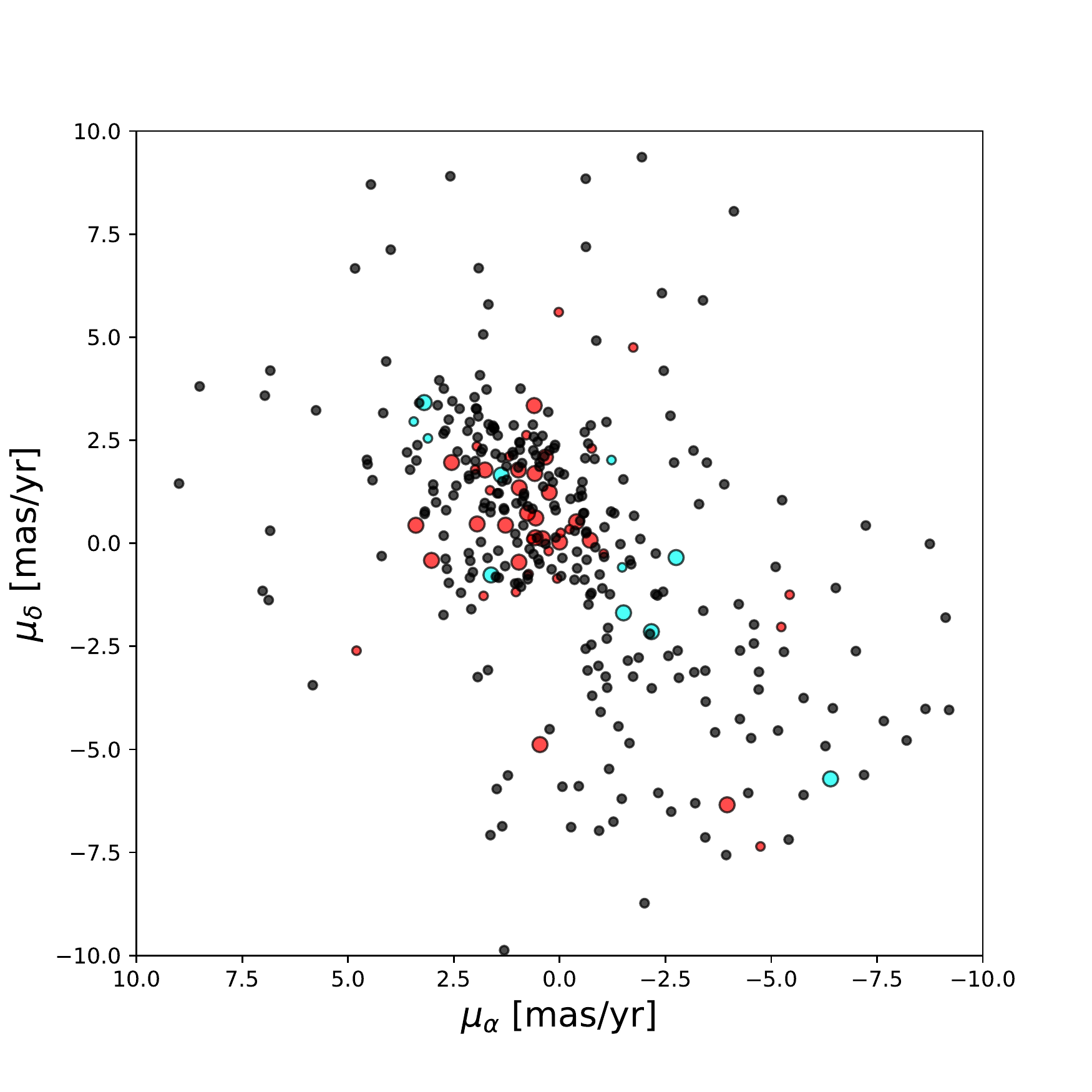}
        }   
        \subfigure{%
            \includegraphics[width=0.48\textwidth]{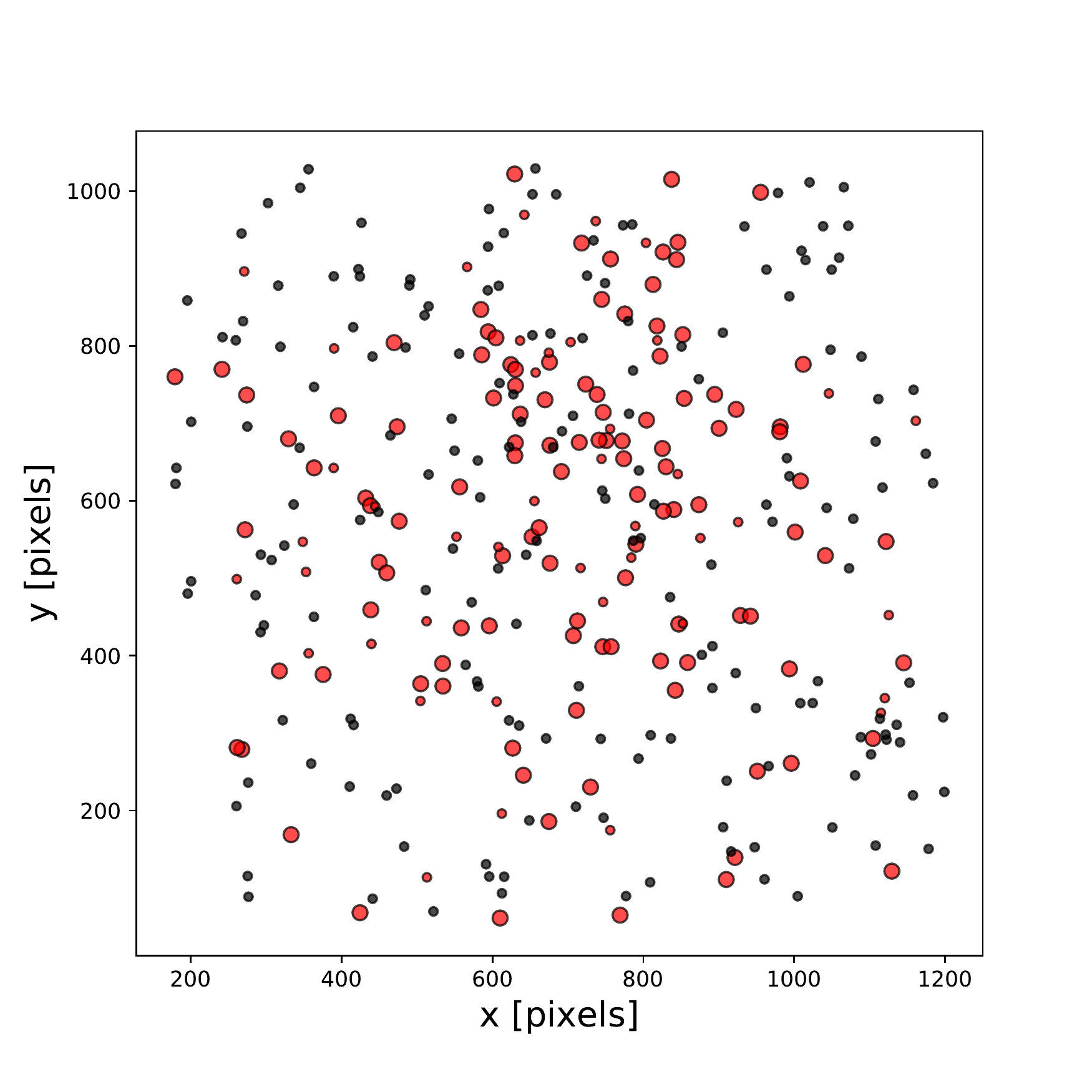}   
        }
        \\

     \end{center}
    
    \caption{Clustering search in our proper motion catalogue. Left column: Cluster selection in the position space (top) and the same sources shown in the velocity space (bottom). The Quintuplet cluster is shown in red and the new group in cyan. Right column: Cluster selection in the velocity space (top) and the same stars shown in the position space in red (bottom). In all panels, the larger coloured points are the core points and the smaller ones are the border points.
    }
\label{fig:dbscan_plots}
\end{figure*}

\begin{figure*}[]
      \begin{center}
    
        \subfigure{
            \includegraphics[width=0.48\textwidth]{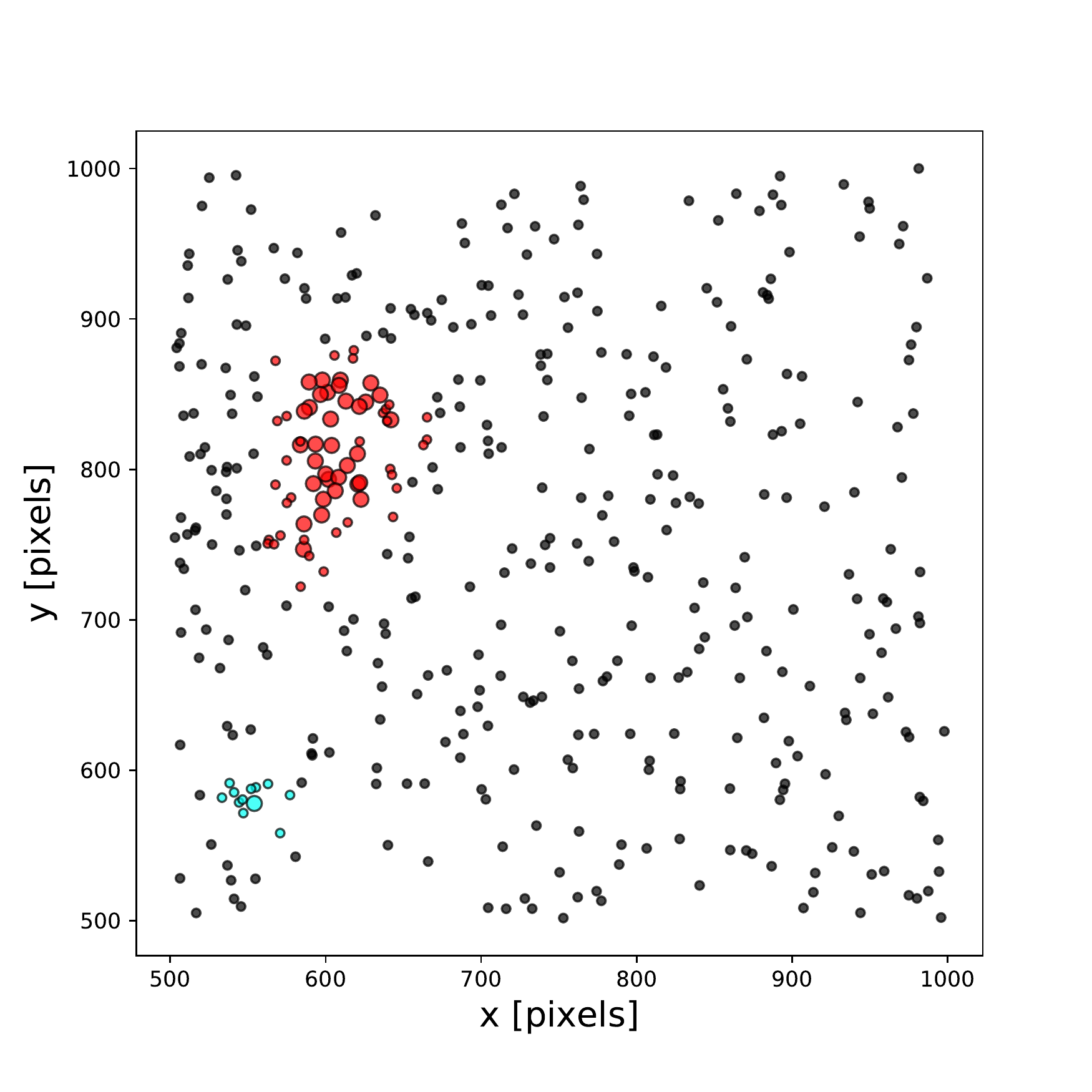}
        }
        \subfigure{%
           \includegraphics[width=0.48\textwidth]{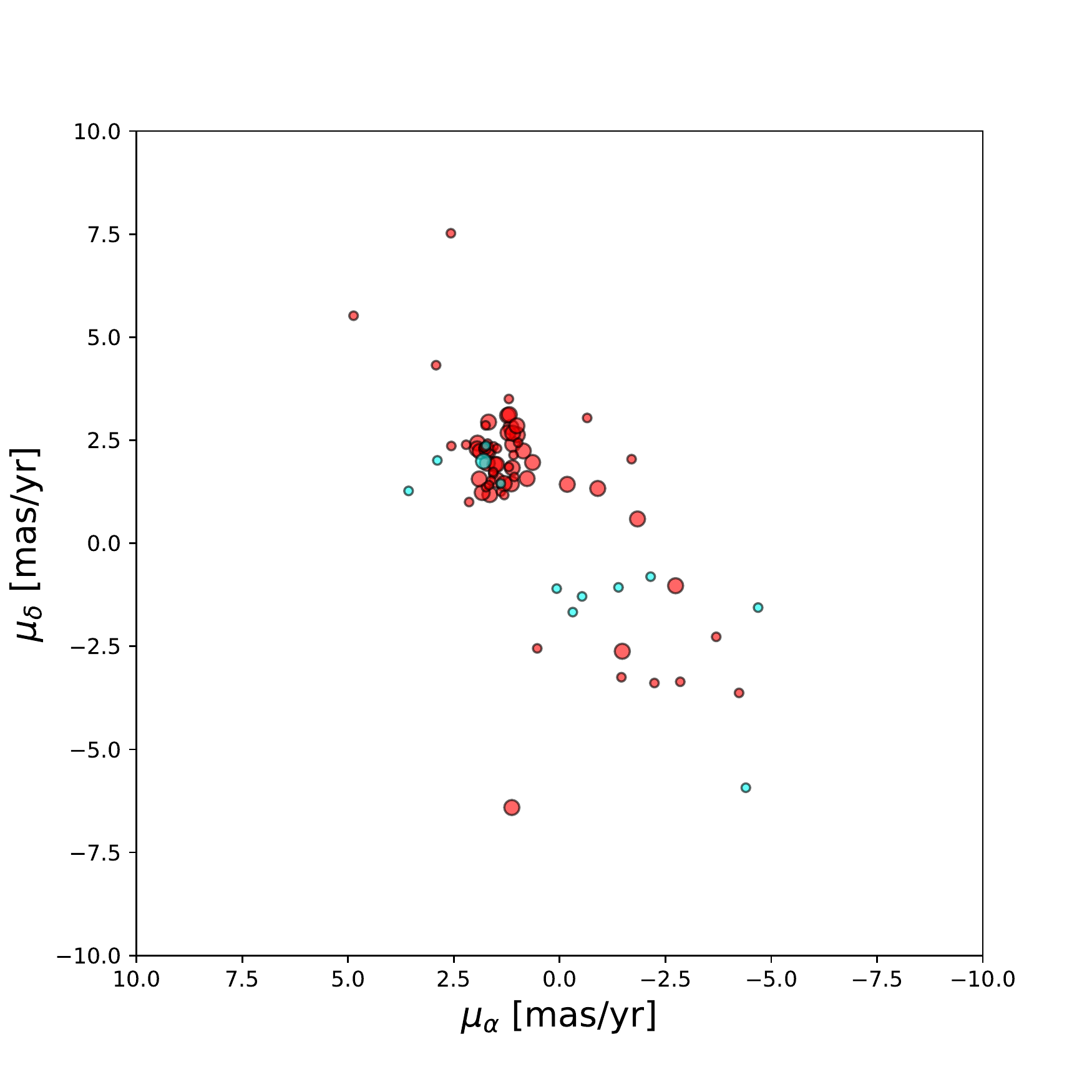}
        }
        \\

     \end{center}
    
    \caption{Clustering search in WFC3/HST data. Left: Cluster selection in the position space. Right: Same sources shown in the velocity space. The Quintuplet cluster is shown in red and the new group in cyan. In all panels, the larger coloured points are the core points and the smaller ones are the border points.}
\label{fig:dbscan_hst}
\end{figure*}

Moreover, we investigated whether we can find this new group of stars in the WFC3/HST data described in Sect.~\ref{section:Comparison with WFC3/HST data}. In Fig.~\ref{fig:quin-hst-overplot} we show the Quintuplet cluster proper motions together with the new co-moving group marked on the WFC3/HST image. Applying the DBSCAN algorithm to these data, we can find the Quintuplet cluster and the new group (see Fig.~\ref{fig:dbscan_hst}). The uncertainties of proper motions in these data are smaller compared to our proper motion catalogue, and as a result there is less scatter in the vector point diagram shown in the panel right of Fig.~\ref{fig:dbscan_hst}.
We see clearly in this panel that five stars of the new group move with the Quintuplet cluster and that the remaining sources in this group are field stars that could be at any distance; therefore, they are not necessarily a group.

\begin{figure*}[]
      \begin{center}
    
        \subfigure{
            \includegraphics[width=0.48\textwidth]{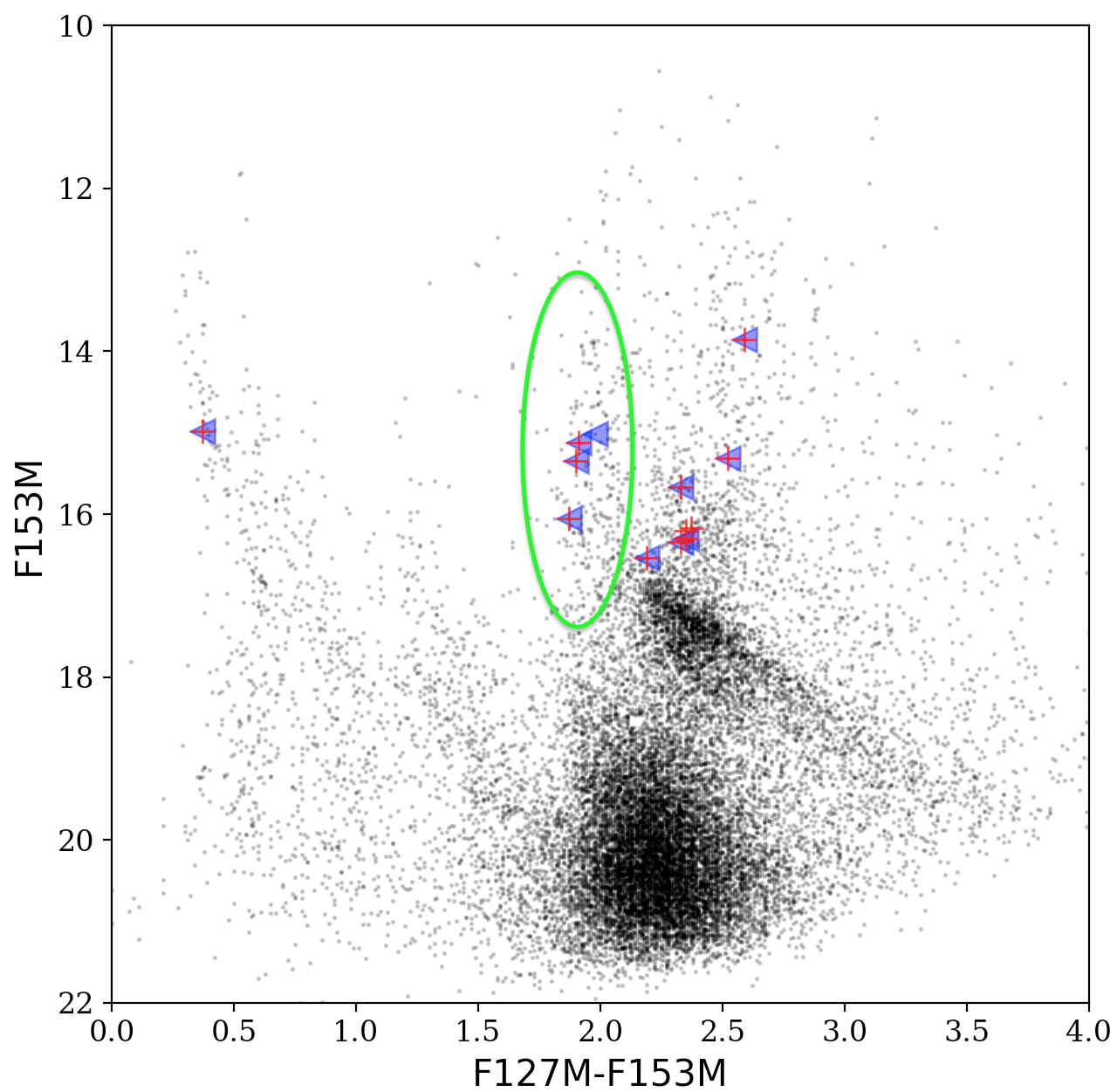}
        }
        \subfigure{%
           \includegraphics[width=0.48\textwidth]{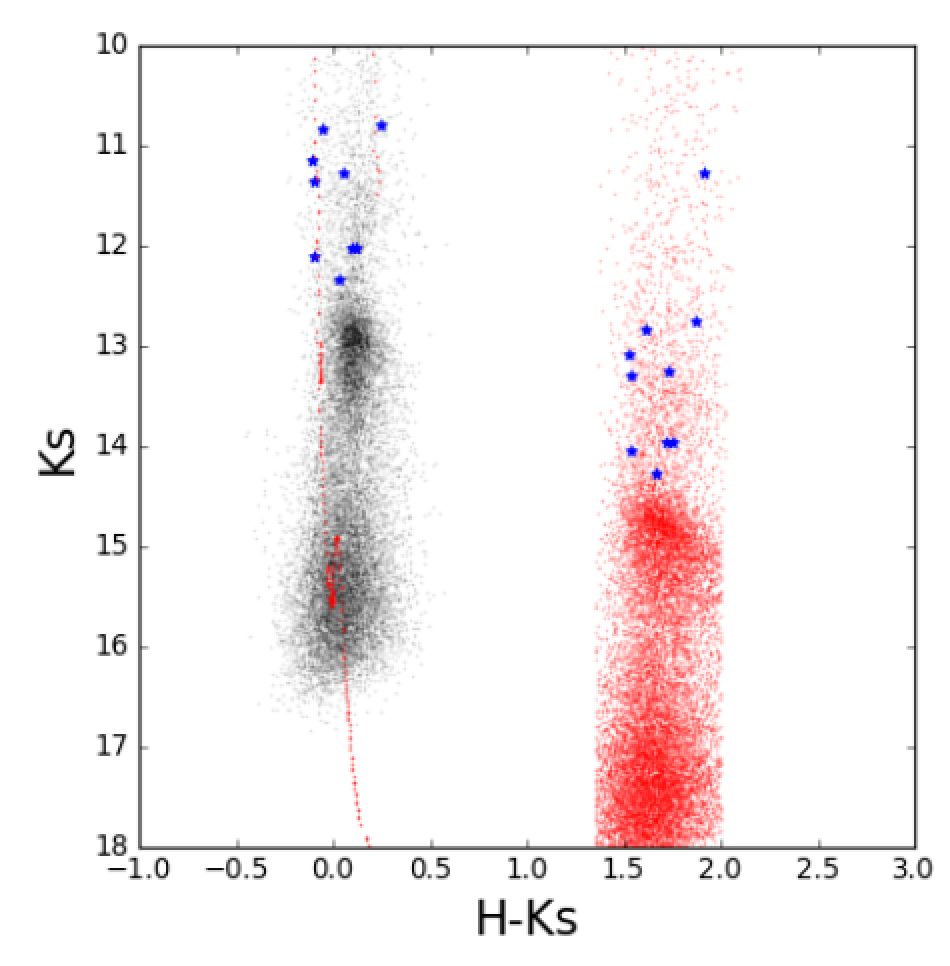}
        }
        \\

     \end{center}
    
    \caption{CMD of the new group. Left: CMD of stars using WFC3/HST data. The red crosses are the new co-moving group sources found in the WFC3/HST data, and the blue triangles are the ones found in our proper motion catalogue. The source with F127M-F153M < 0.5 is a foreground source. The coloured points inside the green ellipse are the new group sources moving in the direction of the Quintuplet cluster movement. Right: CMD of the new group using the GALACTICNUCLEUS data. The co-moving group of stars, excluding the foreground star, is marked in blue. The de-reddened CMD is shown by black points, and an isochrone of 5 Myr old is over-plotted.}
    \label{fig:cmd_new_cluster}
    \end{figure*}

Figure~\ref{fig:cmd_new_cluster} (left) shows the CMD of stars in the region using WFC3/HST data, with the new group found in both these data and our proper motion catalogue indicated.
In this figure the source with F127M-F153M < 0.5 is a foreground source. Comparing this plot with the CMD in Fig.~6 of \cite{rui:2019} shows that the members of the new group in the direction of the Quintuplet cluster (inside the green ellipse) are possibly four stars of the Quintuplet cluster that appear close in the sky. The remaining coloured sources (outside the green ellipse) are probably old giant interlopers that are field stars and do not necessary move coherently. The CMD of the new group using the GALACTICNUCLEUS data is also shown in the right plot of Fig.~\ref{fig:cmd_new_cluster}. We have also produced a de-reddened CMD in this plot using a dedicated extinction map following the methodology described in Appendix A of \cite{nogueras:2021}. The isochrone plotted over the CMD illustrates a 5 Myr old stellar population using PARSEC evolutionary models \citep[release v1.2S + COLIBRI S$\_$35 + PR16][]{Bressan2012,Chen2014, Chen2015, Tang2014, marigo2017, Pastorelli2019}.

Our technique can detect small co-moving groups of stars, and in the future we want to use it across the entire region of our study. Because this approach is density-based and the density might vary greatly from one region to the next, the entire region should be divided into smaller sub-regions for the DBSCAN to be run on each.


\section{Discussion and conclusions}
\label{section:summary}

Stellar kinematics can allow us to disentangle the overlapping and co-penetrating components of the GC (the Galactic disc, bulge or bar, NSD, and NSC) and detect and characterise so far undiscovered young clusters as co-moving groups. This work presents our first step in providing precision proper motion measurements for a large fraction of the GC. We have previously demonstrated our methodology for proper motion measurements on a small GC field in \cite{Shahzamanian:2019} and discovered a new co-moving group of stars.

In this work we have used all overlapping fields between the epoch 2008 H-P$\alpha$S and the epoch 2015-2016 GNS to create the first proper motion catalogue for the central $\sim 36' \times 15'$ of the Milky Way's NSD.  We have made this catalogue, which comprises $\sim$$80,000$ stars (Appendix\,\ref{app:catalogue}), publicly available.

Given that the polynomial fits used for image alignment can diverge at the edges of the images, the small FoV of NICMOS/HST images poses one of the main limitations of our work. The second most important limitation is the low signal-to-noise ratio of the narrow-band HST images, which limits the number of detected sources and their astrometric precision. Due to these limitations, the NICMOS data allow us to measure  the proper motions of only a few percent of the stars detected in GNS, and we can only reach the brightest red clump stars. Nevertheless, this proper motion dataset for the GC is unprecedented in terms of the combination of its number of sources, the quality of the proper motions, and the FoV.  

We clearly detect the presence of at least two GC stellar populations in the kinematic data: stars belonging to the NSD and stars belonging to the bulge. The bulge population has a net zero proper motion and a velocity dispersion of $3-3.17$\,mas\,yr$^{-1}$, in agreement with the literature. Stars in the NSD have a smaller velocity dispersion: $1.5$\,mas\,yr$^{-1}$ in the direction perpendicular to the Galactic plane and about 
    $2-2.5$\,mas\,yr$^{-1}$ parallel to the Galactic plane. The small FoV of the NICMOS images makes precision alignment between the two epochs difficult and impedes us from registering our proper motions in an absolute frame of reference (Gaia). This may give rise to systematic errors that increase the observed velocity dispersions. We therefore caution that the measured velocity dispersions of the nuclear disc population are probably biased towards higher values and should be interpreted as upper limits. 
    
     We clearly detect the rotation of the NSD and the effect of differential reddening along the line-of-sight through the GC. 
    The mean velocities of eastward and westward moving stars correspond to 80\,km\,s$^{-1}$, in agreement with literature values derived from line-of-sight velocities of stars in the nuclear disc.

We compared our NSD velocity distributions to those obtained by axisymmetric self-consistent dynamical models of the NSD from \cite{Sormani2021}. Their models are based on kinematics and do not take photometric information into account. We selected the NSD stars from the same region as our study (see Fig.~1) from their model-generated data and fitted Gaussians with the \textit{dynesty} package. As a result, we obtain a velocity dispersion of $1.49$\,mas\,yr$^{-1}$ for stars in the direction perpendicular to the Galactic plane, which agrees well with our observations (see Table~\ref{table:data}). However, the velocity dispersion obtained for the stars in the direction parallel to the Galactic plane is about $1.5$\,mas\,yr$^{-1}$, which is less than what we find from our data. The larger velocity dispersion of the nuclear disc that we get for $\mu_{l}$ compared to \cite{Sormani2021} models is likely due to systematic uncertainties that we have because of the small FoV of NICMOS data, which prevents us from registering our proper motions in an absolute reference frame.

Furthermore, we demonstrate a technique, based on a density-based clustering algorithm, that shows how proper motions can be used to identify new potential young co-moving groups, which can be subsequently targeted with instruments with smaller FoVs but higher angular resolution for more detailed follow-up studies. We applied the method to the Quintuplet cluster region, which resulted in the detection of a new potential co-moving group of stars. However, further analysis shows that four sources of this group are possibly stars of the Quintuplet cluster and that the remaining sources are field stars.

In future work, we will expand this catalogue with a new HAWK-I imaging epoch. The higher sensitivity and larger FoV of HAWK-I mean that we will be able to measure the kinematics of about 100 times more stars, minimise the uncertainties of aligning data from different epochs, and tie the measured proper motions into the Gaia reference frame. We will then obtain a much clearer view of the kinematics of the  stellar populations at the GC.


\begin{acknowledgements}

The authors would like to thank the anonymous referee
for the helpful comments on this paper. B.Sh, R.S, and A.T.G.C acknowledge financial support from the State Agency for Research of the Spanish MCIU through the ''Center of Excellence Severo Ocho'' award for the Instituto de Astrof\'isica de Andaluc\'ia (SEV-2017-0709). B.Sh, A.T.G.C, A.A, and R.S acknowledge financial support from national project PGC2018-095049-B-C21 (MCIU/AEI/FEDER, UE). F.N.L and M.C.S gratefully acknowledge support by the Deutsche Forschungsgemeinschaft (DFG, German Research Foundation) Project-ID 138713538 SFB\,881 (The Milky Way System, subproject B8). M.C.S furthermore acknowledges support from the ERC via the ERC Synergy Grant “ECOGAL” (grant 855130). F.N.L acknowledges the sponsorship provided by the Federal Ministry for Education and Research of Germany through the Alexander von Humboldt Foundation. This work is based on observations made with ESO Telescopes at the La Silla Paranal Observatory under programmes IDs 195.B-0283. We thank the staff of ESO for their great efforts and helpfulness. This research is based on observations made with the NASA/ESA Hubble Space Telescope

 obtained from the Space Telescope Science Institute, which is operated by the Association of Universities for Research in Astronomy, Inc., under NASA contract NAS 5–26555. These observations are associated with programs 11671, 12318, 12667, and 14613.

\end{acknowledgements}

\bibliographystyle{aa} 
\bibliography{pm_catalogue_paper} 

\newpage

 \begin{appendix}
  \onecolumn 
 \section{Description of the proper motion catalogue \label{app:catalogue}}
 
We obtained proper motions for GC stars in the near-infrared H band for 77414 stars. The first lines of the proper motion catalogue are shown in Table~\ref{tab:data_cat}.

 \begin{table*}[htbp]
 \caption{First rows of the proper motion catalogue. }
        \begin{center}
        \begin{tabular}{cccccccccc}
        \hline\hline
        \\RA  & $\Delta$RA & Dec & $\Delta$Dec & H & dH & $\mu_{l}$ & $d\mu_{l}$ & $\mu_{b}$ & $d\mu_{b}$ \\ 
        (deg) & (arcsec) & (deg) & (arcsec) & (mag) & (mag) & (mas\,yr$^{-1}$) & (mas\,yr$^{-1}$) & (mas\,yr$^{-1}$) & (mas\,yr$^{-1}$)\\
        \\ \hline \\[0.1ex]
        266.41925 &0.00097& -28.80355& 0.00068& 15.21485& 0.01091& -4.19960& 0.94585& -0.21576& 0.89140\\
        266.38788 &0.00126& -28.79402& 0.00154& 14.09413& 0.01906& 1.51912& 1.19169& -0.25583& 1.44118\\
        266.39636 &0.00139& -28.78592& 0.00111& 16.20390& 0.01702& -2.75357& 1.07579& 5.50186& 1.34625\\
        266.46149 &0.00024& -28.82738& 0.00028& 15.90056& 0.00269& 3.25698& 1.23304& -3.95698& 0.80972\\
        266.45837 &0.00092& -28.83129& 0.00112& 16.74176& 0.00426& -1.01231& 1.78124& 0.70804& 0.72085\\
        266.45883 &0.00035& -28.81885& 0.00042& 12.33349& 0.00333& -0.38286& 0.67396& -5.06638& 0.57996\\
        266.46439 &0.00033& -28.82380& 0.00042& 13.23098& 0.00451& 0.15466& 0.63118& -0.47885& 0.66669\\
        266.46310 &0.00017& -28.82315& 0.00027& 13.50527& 0.00324& 0.81493& 0.59243& 1.39099& 0.55884\\
        266.45731 &0.00042& -28.81703& 0.00035& 13.86630& 0.00316& 6.76527& 0.75450& 0.18626& 0.71306\\
        266.46158 &0.00087& -28.82217& 0.00100& 14.26329& 0.01246& 1.26582& 0.63530& 2.36740& 0.55892\\
        266.45984 &0.00026& -28.82113& 0.00040& 14.27947& 0.00237& 0.73303& 0.57984& 0.71113& 0.56797\\
        266.46005 &0.00046& -28.82115& 0.00046& 14.41241& 0.00353& 1.00725& 0.58351& -1.34299& 0.69243\\
        266.46002 &0.00030& -28.82071& 0.00031& 14.60487& 0.00341& 1.09343& 0.58009& 2.26018& 0.52826\\
        266.46204 &0.00034& -28.82129& 0.00042& 14.71005& 0.00485& -2.38172& 0.61363& -0.01849& 0.56018\\
        266.46249 &0.00035& -28.82142& 0.00042& 14.88699& 0.00399& -0.97400& 0.59609& 0.30383& 0.56781\\
        266.45737 &0.00020& -28.81890& 0.00020& 14.89931& 0.00402& -5.31594& 0.66412& 2.53803& 0.61431\\
        266.46188 &0.00042& -28.82163& 0.00035& 14.89672& 0.00324& -0.38727& 0.61505& -0.00836& 0.77331\\
        266.46045 &0.00042& -28.82105& 0.00035& 15.22904& 0.00245& 2.62466& 0.58435& 1.66795& 0.53053\\
        266.46039 &0.00058& -28.82126& 0.00050& 15.43128& 0.00509& 0.09456& 0.58066& 3.10825& 0.62786\\
        266.46262 &0.00053& -28.82263& 0.00049& 16.10555& 0.00687& 1.81359& 0.64299& 4.89179& 0.56042\\
        266.46057 &0.00061& -28.82119& 0.00061& 16.11472& 0.00427& 3.45930& 0.60339& 2.09749& 0.61913\\
        266.46198 &0.00026& -28.82258& 0.00023& 16.15639& 0.00477& 1.15596& 0.58290& 0.23051& 0.61128\\
        266.45825 &0.00067& -28.81790& 0.00047& 16.35515& 0.00604& -5.25076& 0.70984& -1.35620& 0.67907\\
        266.46320 &0.00056& -28.82372& 0.00059& 16.37724& 0.00444& -5.08730& 0.65269& -0.07486& 0.58265\\
        266.45694 &0.00047& -28.81782& 0.00067& 16.44830& 0.00435& 5.46795& 0.76426& 2.57502& 0.77120\\
        266.46384 &0.00080& -28.82331& 0.00069& 16.48845& 0.00920& -0.67524& 0.65292& 1.26968& 0.60490\\
        266.46027 &0.00096& -28.82076& 0.00103& 16.48476& 0.03141& -0.32617& 0.67995& 1.69349& 0.59741\\
        266.46173 &0.00032& -28.82399& 0.00031& 12.03400& 0.00488& -0.82098& 0.50486& -1.43832& 0.47763\\
        266.45947 &0.00053& -28.82145& 0.00034& 12.44773& 0.00979& 1.21442& 0.51689& 0.07001& 0.46946\\
        266.46021 &0.00033& -28.82553& 0.00031& 12.48363& 0.00448& 0.74789& 0.45218& -0.35204& 0.41850\\
        266.46112 &0.00030& -28.82294& 0.00022& 12.56305& 0.00632& -1.28993& 0.50806& 0.88989& 0.42835\\
        266.45703 &0.00016& -28.82382& 0.00017& 12.91874& 0.00407& -0.64706& 0.45378& 0.11141& 0.40211\\
        266.45789 &0.00040& -28.82413& 0.00027& 13.08759& 0.00242& 2.94974& 0.43846& -1.99291& 0.43946\\[0.1ex]
        ...  & ... &...  &...  &...  &...  &...  &...  &...  & ...\\

        \hline
        \end{tabular}   
        \end{center}
        \begin{tablenotes}
        \small
        RA, $\Delta$RA and Dec, $\Delta$Dec are the right ascension and the declination of the sources (and their corresponding uncertainties). H and dH are the H-band magnitude and its uncertainty. $\mu_{l}$, $d\mu_{l}$, and $\mu_{b}$, $d\mu_{b}$ are the proper motions in the direction parallel and perpendicular to the Galactic plane (and their corresponding uncertainties).
        
        \end{tablenotes}
        \label{tab:data_cat}
       
        \end{table*}

\end{appendix}
\end{document}